\documentclass[useAMS,usenatbib]{mn2e}
\usepackage{graphicx}
\usepackage[latin1]{inputenc}
\usepackage[portuges]{babel}
\usepackage{subfigure}
\usepackage{indentfirst}
\usepackage{picinpar}
\usepackage{multicol}

\title[Stability Regions Around the Components of the Triple System 2001 SN263]
{Stability Regions Around the Components of the Triple System 2001 SN263}
\author[R.A.N.Araujo et al.]{R.A.N.Araujo$^{1,2}$\thanks{E-mail:
ran.araujo@gmail.com},O.C.Winter$^{1,2}$ \thanks{E-mail:
ocwinter@pq.cnpq.br}, A.F.B.A.Prado$^{1}$ \thanks{E-mail:
prado@dem.inpe.br} and A. Sukhanov $^{1}$ \thanks{E-mail:sasha.su@hotmail.com} \\
$^{1}$INPE - National Institute for Space Research, CEP 12201-970 - S\~ao Jos\'e dos Campos, SP, Brazil.\\
$^{2}$UNESP - S\~ao Paulo State University,CEP 12516-410, Guaratinguet\'a, SP, Brazil.\\}

\begin{document}

\date{Accepted for publication in 2012 April 4.}

\pagerange{\pageref{firstpage}--\pageref{lastpage}} \pubyear{2012}
\maketitle
\label{firstpage}

\begin{abstract}
The space missions are an unquestionable way to increase our knowledge about asteroids. The NEAs (Near-Earth Asteroids) arise as good targets 
for such missions, since they periodically approach the orbit of the Earth. Due to these advantages, a growing number of missions to NEAs are being
 planned around the world. Recently, the NEA (153591) 2001 SN263 was chosen as the target of the ASTER MISSION- First Brazilian 
Deep Space Mission, planned to be launched in 2015. The NEA (153591) 2001 SN263 was discovered in 2001. In February 2008, the radio astronomers from Arecibo-Puerto 
Rico concluded that (153591) 2001 SN263 is actually a triple system \citep{b6}. The announcement of the ASTER MISSION has motivated the development of the present
 work, whose goal is to characterize regions of stability and instability of the triple system (153591) 2001 SN263. Understanding and characterizing the stability 
of such system is an important information to design the mission aimed to explore it. The method adopted consisted in dividing 
the region around the system into four distinct regions (three of them internal to the system and one external). We have performed numerical integrations of systems composed by seven bodies: Sun, Earth, Mars, 
Jupiter and the three components of the system (being Alpha the most massive body, Beta the second most massive body, and Gamma, the least massive body), 
and by thousands of particles randomly distributed within the demarcated regions, for the planar and inclined prograde cases. The results are diagrams of 
semi-major axis versus eccentricity, where it is shown the percentage of particles that survive for each set of initial conditions. 
The regions where $100\%$ of the particles survive is defined as stable regions. We found that the stable regions are in the neighborhood of Alpha and 
Beta, and in the external region. It was identified resonant motion of the particles with Beta and Gamma in the internal regions, which lead to instability. 
 For particles with $I>45^{\circ}$ in the internal region, where $I$ is the inclination with respect to Alpha's equator, there is no stable region, except for the particles placed really close to Alpha. 
The stability in the external region is not affected by the variation of inclination. We also present a discussion on the long-term stability in the 
internal region, for the planar and circular cases, with comparisons with the short-term stability.

\end{abstract}


\section{Introduction}

Asteroids are bodies that orbit the Sun but are too small to be considered planets and, as examples of primordial objects, they can help us to understand the process 
of formation of our Solar System. They may receive different classifications according to their orbital characteristics, and according to their physical, chemical 
and mineralogical properties. An interesting group of asteroids is formed by the NEAs (Near-Earth Asteroids) and as the name suggests, designates asteroids that 
approach Earth's orbit periodically.

The NEAs belong to the group of NEOs (Near Earth Objects), and updated data from NASA show that they represent about $99\%$ 
of such population. The NEAs are classified into four groups, according to their orbital characteristics given by perihelion (q), by aphelion (Q) and the 
semi-major axis (a), compared with the orbital characteristics of the Earth: Aphelion $ Q_ {T} = 1.017$ AU, perihelion $ q_ {T} = 0.983$ AU and semi-major axis $ a_{T} = 1.0$ AU.
They are: Apollo (corresponding to about $54.5\% $ of the population of NEOs with $q<Q_{T}$, $a>a_{T}$ and being an Earth-crossing),  Aten (corresponding to about $8.3\% $ of the 
population with $Q>q_{T}$, $a<a_{T}$ and also an Earth-crossing),  Amor (about $37.1\%$ of the population with $Q_{T}<q<1.3$ AU., $a>a_{T}$. They cross the orbit of Mars and
approach the orbit of the Earth without crossing it), and the Atira group, or IEO - Interior to the Earth Orbit (about $0.1\%$ of the population with $Q<q_{T}$, $a<a_{T}$.
As the name suggests, they are NEOs  with orbits internal to the orbit of Earth without crossing orbits).

An interesting subgroup of asteroids is formed by the binary and multiple systems. The first triple system of main belt asteroids discovered was 87 Sylvia in the main belt \citep{b19}. 
The study of the dynamics of such system was done by \cite{b20}. 

Examination of doublet craters on Earth and Venus led  \cite{b17} to suggest that about $15\%$ of the asteroids that cross the orbit of Earth are binaries.
This suggestion was verified observationally by \cite{b23} and \cite{b24}. For Mars-crossing asteroids the expected number is $5\%$. Currently, $36$ asteroids were identified as multiple system on the group of NEAs, being $34$ of them binary 
system, and only two triple systems: (153591) 2001 SN263 (Amor) \citep{b6} and 1994 CC (Apollo) \citep{b25}, discovered as a triple system in 2008 and 2009 respectively. 
Hereafter, we are going to refer to the triple system (153591) 2001 SN263 just by its provisional designation: 2001 SN263.

The NEOs are celestial bodies that represent a threat to life on Earth. Due to the possibility of a catastrophic event caused by impacts of such bodies with the surface
of our planet, and thus, besides the effort to increase the number of known NEOs is also important to know the composition of these bodies, to understand the process that 
gives rise to them, and examine how they evolve dynamically. This have been done over the last few decades and as examples we can mention 
\cite{b11} with a study on the lifetime of NEOs (about a few million years).  \cite{b12} presented the first study aimed to explain the mechanism responsible for maintaining 
the population of NEOs, indicating a cometary origin for them.  Later, the asteroidal origin of the NEOs has been considered, as discussed in \cite{b14} which concluded that some NEOs with high values of eccentricities and inclination would be extinct nuclei of comets
while the remaining NEOs, known at that time, were main belt asteroids and that they reached typical NEOs orbits by encounters with the planet Mars. In $2002$, \cite{b15}
showed how asteroids can escape from the asteroid belt through resonance mechanisms and supply the known population of NEOs. According to their study only $6\%$ of the NEOs
population would come from the Kuiper Belt (cometary origin). 

Besides the theoretical studies, as the exemplified above,  the space missions are another unquestionable way to increase our knowledge about asteroids. 
The NEAs arise as good targets for such missions, since they periodically approach the orbit of the Earth.  In this scenario, 
the NEAs composed by two or three bodies (multiple asteroid systems) are especially interesting since they increase the range of possible observations and scientific results 
obtained by the mission. Due to these advantages a growing number of missions to NEAs have been completed or are being planned by the major aerospace agencies with these
body as target. 

In May 2003, the Japan Aerospace Agency launched the Hayabusa mission to the NEA (25413) Itokawa, which reached the target in September 2005. The same agency plans a new mission 
called Hayabusa 2, targeting the NEA 1999 JU3, applying the same technology and concepts of the first mission \citep{b10}.  The European Space Agency (ESA) has some 
studies about missions to NEAs. The program Don Quijote \citep{b5} planned to launch two spatial vehicles. One of them would collide with some NEA and the second one 
would catch information about the internal structure of the asteroid. The program ISHTAR \citep{b2} was planned to visit at least two NEAs and characterize all physical 
parameters of the asteroids, such as its mass distribution, density and surface properties. The SIMONE mission \citep{b9} was planned to be composed of five micro-satellites which will study individually NEAs of different classifications. The Marco Polo mission is also part of the program of ESA missions, and its main objective is to return to Earth carrying a sample of a NEA \citep{b1}. The HERA mission is a project being developed by Arkansas Center for Planetary Science and the Jet Propulsion Laboratory. The goal is to send a probe to collect samples for 3 NEAs and then return to the Earth \citep{b7}.

Recently, the NEA 2001 SN263 was chosen as the target of the ASTER MISSION- First Brazilian Deep Space Mission, planned to be launched
in 2015 \citep{b8}. The target was chosen taking into account the advantages of send a spatial probe to a multiple system of asteroids, which increases the range of possible 
scientific investigation (such as internal structure, formation process, and dynamical evolution), with the economy of fuel, flight time and telecommunication system
required, when compared to a similar mission aimed to an asteroid of the main belt. The announcement of such project has motivated the development of the present work, whose 
goal is to characterize the regions of stability and instability of the system.  Such information is of great interest since it will serve as a parameter 
for the mission planning, and there are no previous work already done in this direction.
The study was made through numerical integration of the system composed by seven massive bodies
(Sun, Earth, Mars, Jupiter and the components of the triple system), and by thousands of particles randomly distributed around the three asteroids (internal regions) and around the whole system 
(external region).

The structure of this paper is such that, in section 2, we present the triple system 2001 SN263. In section 3, we discuss the initial conditions of the problem
and the methodology adopted. In section 4, we present the study on the stability in the internal region of the system for the planar case. 
In section 5, we present similar study considering the inclined prograde case. In section 6, we discuss the resonant motion identified in the internal region.
In section 7, we present the study on the stability in the external region. In section 8, we present the analysis on the long-term stability in the internal region, 
for the planar and circular case. In section 9, we present the final comments with an overview of the results presented in the previous sections.

\section{The triple system of asteroid 2001 SN263}
\label{sec_system}

The asteroid 2001 SN263 was discovered in 2001 by the program LINEAR (Lincoln Near-Earth Asteroid Research) - a program developed jointly by the U.S. Air Force,
 NASA and the Lincoln Laboratory. Light-curves obtained in the Observatory of Haute-Provence, in January 2008, lead to the conclusion that this asteroid was a binary system.
 In February 2008 the system was observed along 16 days by the radio-astronomy station of Arecibo, in Puerto Rico. Those observations led to the discovery that 
2001 SN263 is a triple system \citep{b6}.  This is the first triple system known that approaches the orbit of the Earth and that crosses the orbit of Mars 
(Amor type asteroid). In January 2009, \citep{b16} presented preliminary data on the physical aspects of the asteroids. Their study estimated that the 
primary body (largest) is approximately a spheroid with principal axes of approximately $2.8 \pm 0.1 km$, $2.7\pm 0.1 km$, $2.5\pm  0.2 km$, with an estimated density 
of $1.3\pm0.6$ $g/cm^{3}$.  

In $2011$, \cite{b4} presented a dynamical solution for both triple system known on the group of NEAs: 2001 SN263 and 1994 CC. 
Using the data obtained through radar observation from Arecibo and Goldstone, and through numerical integrations
of the N-body problem, they derived the masses of the components, the $J_{2}$ gravitational harmonic of the central bodies, and orbital parameters of the satellites.
The orbital and physical data for 2001 SN263 can be seen in Table $1$.

\begin{figure}
\begin{center}
\includegraphics[scale=0.044]{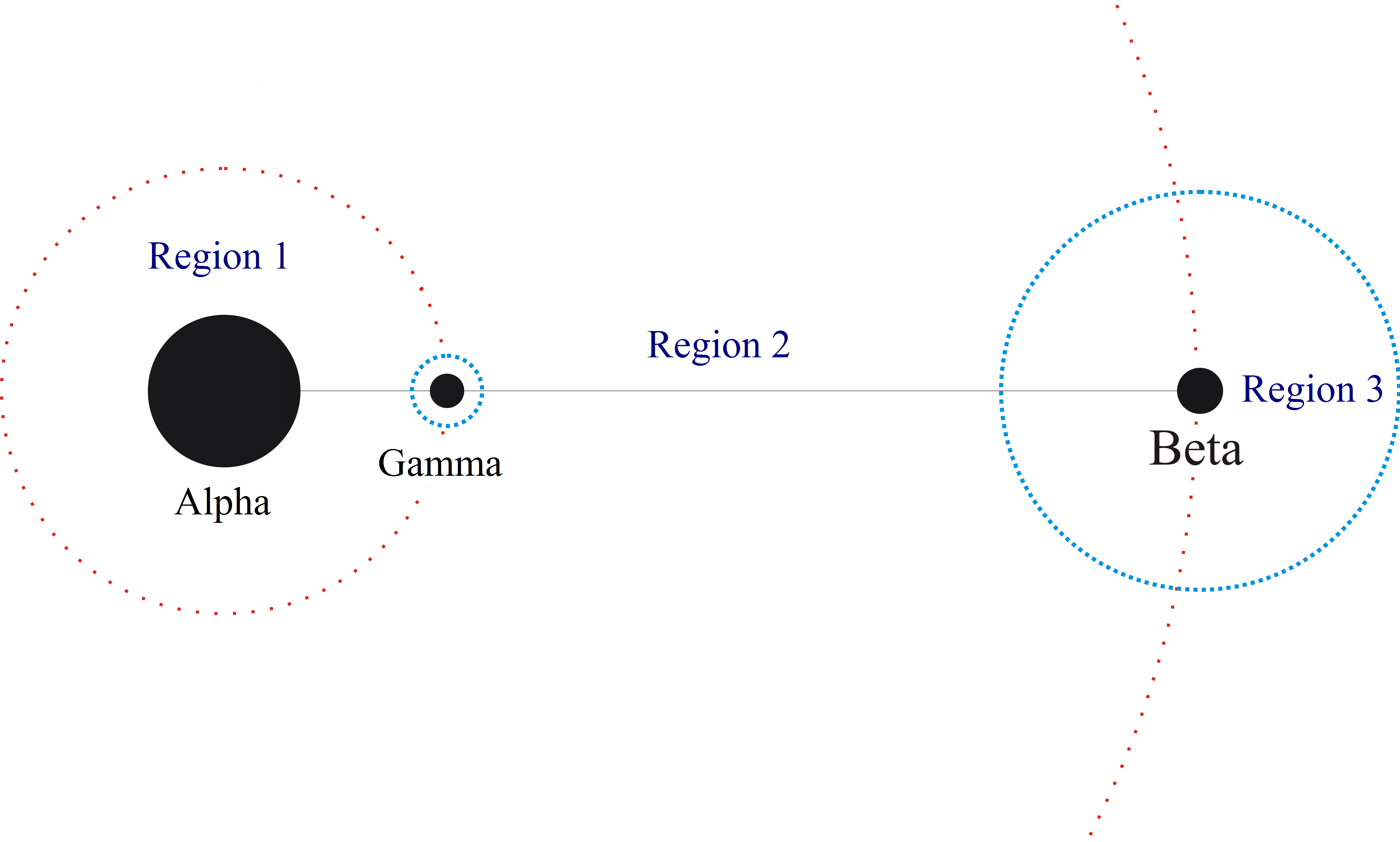}
\end{center}
\caption{Representation of the triple system 2001 SN263. The blue circles represent the Hill's radii of Beta and Gamma. 
The red dotted circles represent the collision-lines with Gamma and Beta, and by definition, the limits of the internal regions.}
\label{fig_systemhill}
\end{figure}

Hereafter, we follow the nomenclature adopted by \cite{b4} for the triple system. We refer to the central body (most massive body) as $Alpha$, the second most massive body is called $Beta$ (outer)
and the least massive body is called $Gamma$ (inner). Figure \ref{fig_systemhill} is a representation of the system 2001 SN263.

\begin{table*}
\centering
\begin{minipage}{120mm}
\centering
\caption{Physical and orbital data of the three components of the system 2001 SN263}
\end{minipage}
\begin{tabular}{@{}ccccc @{}}
\begin{tabular}{ |c|c|c|c|c|c|c|c| }
\hline
Body	&Orbits   &a$^{1}$	 &e$^{1}$	 &I$^{1}$$^{(*)}$     &Period$^{1}$	&Radius     		&Mass$^{1}$ \\
\hline
Alpha		&Sun	  &$1.99$ AU    &$0.48$ 	 &$6.7^{\circ}$      &$2.80$ years      &$1.3$ km $^{1} $       &$M_{\alpha}=917.47\times10^{10}$ kg \\
\hline
Beta		&Alpha  &$16.63$ km     &$0.015$   	 & $0.0^{\circ}$     &$6.23$ days       &$0.39$ km$^{(**)}$               &$M_{\beta}=24.04\times10^{10}$ kg \\
\hline
Gamma		&Alpha  &$3.80$ km     &$0.016$ 	 &$\approx14^{\circ}$   &$ 0.69$ days	&$0.29$ km$^{(**)}$            &$M_{\gamma}=9.77\times10^{10}$ kg \\
\hline
\multicolumn{8}{l}{ 1 - \cite{b4}.}\\
\multicolumn{8}{l}{(*) Inclination of Alpha related to the ecliptic plane. Inclinations of  Beta and Gamma related to the}\\
\multicolumn{8}{l}{equator of Alpha.}\\
\multicolumn{8}{l}{(**) Estimated values. From the mass and the radius of Alpha we calculated its density. Assuming that}\\
\multicolumn{8}{l}{Beta and Gamma have the same density as Alpha, and knowing their masses, we estimated their radius.}\\
\label{tab_elements}
\end{tabular}
\end{tabular}
\end{table*}

\section{Regions of stability}
\label{sec_methodology}

The triple system 2001 SN263 is an interesting system. It is composed by bodies of similar values of mass and radius, and they are close to each other. A
particle placed on their neighborhood must suffers perturbations coming from all the bodies and this certainly will be decisive in determining the regions of stability, 
and instability, inside and outside of the system.

The determination of such regions would be an indicative of the location of some possible additional component of the system
that was not observed yet or, where small debris would be placed (stable region). On the other hand the unstable region implies that such region is expected to be empty 
(no bodies or debris). Such information is useful for the planning of a mission for the system, as the Brazilian mission cited previously. 

We developed a method to determine regions of stability on the neighborhood of the components of the triple asteroid system
2001 SN263, in terms of orbital elements, within a given time span, considering only the gravitational perturbation. At this first approach the effect due to the radiation 
pressure was not considered, although it is known that it may cause the removal of some of the small particles from the stable regions.

The initial conditions and the methodology adopted are described in the next subsections.

\subsection{Initial Conditions}
\label{sec_initial}

We consider a system composed by seven massive bodies: the three components of the system 2001 SN263, the Sun and the planets Earth and Mars (due to possible close encounters) 
and Jupiter (the system crosses the main asteroid belt).
We introduced on the system thousands of particles, randomly distributed around the three asteroids as follow:

\begin{itemize}
\item{Spatial distribution:  the region around the three bodies was divided into four regions. We calculated the Hill's radius for the 
problems composed by Alpha-Beta and Alpha-Gamma separately. This is an approximation since the presence of the third body will change the values found; however,
the Hill's radius is a good parameter to spatially delimit the regions where each of the bodies is gravitationally dominant. The values found 
were  $R_{Hill}\approx3.4$ km  for the primary bodies Alpha-Beta and $R_{Hill}\approx0.6$ km for the primary bodies Alpha-Gamma. Figure \ref{fig_systemhill} is a 
representation of the asteroid system and of the Hill's radius found for each body. The particles are spatially distributed into those regions, being: region 1, the region 
between Alpha and Gamma, with particles orbiting Alpha. Region 2, the region between Gamma and Beta, with particles orbiting Alpha. Region 3, the region where particles 
are orbiting Beta, limited by the Hill's radius of this body, and , region 4, the region around Gamma, limited by the Hill's radius of this body, and with particles orbiting Gamma.

\item{ All particles start with circular orbits $(e=0.0)$ until eccentricities equal to $0.5$.}

\item{At first, only the planar case was considered (all particles with inclination $I=0.0^{\circ}$ relative to the equator of the central body - Alpha). 
After, we performed the same analysis for particles with inclinations going from $I=15,0^{\circ}$} until $I=90,0^{\circ}$ (prograde cases)}.

\item{Angularly the particles were distributed with random values of true anomaly $(0^{\circ} \leq f \leq 360^{\circ})$,
 argument of pericentre $(0^{\circ} \leq \omega \leq 360^{\circ})$ and longitude of the ascending node $(0^{\circ} \leq \Omega \leq 360^{\circ})$.}

\item{We considered the oblateness of the central body Alpha} with a value of $J_{2}=0.013.$ \citep{b4}. The obliquity of Alpha was also considered. It was determined through
the pole solution given by Fang et al (2011).
 \end{itemize}


\subsection{The Method}
\label{sec_method}

The method adopted is the numerical integration of the equations of motion of the problem composed by seven bodies 
(Sun, Mars, Earth, Jupiter and the triple system 2001 SN263) and by $n-particles$ (the number of particles will change for each region), for a time span of 2 years.

The orbits of the particles follow the conditions explained in the previous subsection. 
The orbital elements of Beta and Gamma used as  initial conditions in the integrations are given  by \cite{b4}.
The data presented by them correspond to the epoch MJD 54509 in the equatorial frame of J2000.  
The orbital elements of the other bodies (Alpha, Earth, Mars and Jupiter) were obtained through the JPL's Horizons 
system for the same epoch.

The time span of 2 years corresponds to $\approx100$ orbital
periods of Beta, and  $\approx1000$ orbital periods of Gamma, and it is sufficient to guarantee the applicability of the results in the planning of the spatial mission 
to the system. The numerical integrations were performed with the Gauss-Radau numerical integrator \citep{b3}.
The timestep of the numerical integrations, for the internal and external regions, were of 1 hour ($\approx 6\%$ of Gamma's orbital period).

Throughout the integration period we monitored the particles that collide with any of the bodies, and the particles ejected from the system.  The collisions depend 
on the radius of the bodies, and the ejection is defined for each region. For region 1, the ejection distance is $d>3.804$ km, which is an approximation, and corresponds to the circular
orbit of Gamma. Similarly, for region 2, the ejection distance is $d>16.633$ km, which is also an approximation, and corresponds to the circular orbit of Beta. The red dotted lines in Figure \ref{fig_systemhill}
represent such ejection distances, and so, the limits of regions 1 and 2. For regions 3 and 4, the ejection distances are defined by the Hill's radius of Beta and Gamma, respectively.
So, $d>3.4$ km for region 3, and, $d>0.6$ km for region 4.

The percentage of collisions and ejections define the regions of stability and instability of the system. The region of stability is defined as the region where $100\%$ of the
particles survive for $2$ years. Below this value, we have instability.

\section{Planar Case}
\label{sec_results}

Here we present the results for each of the regions described in subsection \ref{sec_initial}, considering only particles with $I=0.0^{\circ}$ relative to
the equator of Alpha. 

The region $4$ was not considered on the integrations. It is a very small region ($\approx0.6$ km) and 
would became even smaller considering the gravitational influence due to the third body (Beta). The particles would orbit very close to Gamma
increasing the collisions probability. 

Therefore, the internal regions considered were the regions 1, 2 and 3.


\subsection{Region 1}
\label{subsec_stabregion1}

\begin{figure*}
\mbox{%
\subfigure[]{\includegraphics[height=7.5cm]{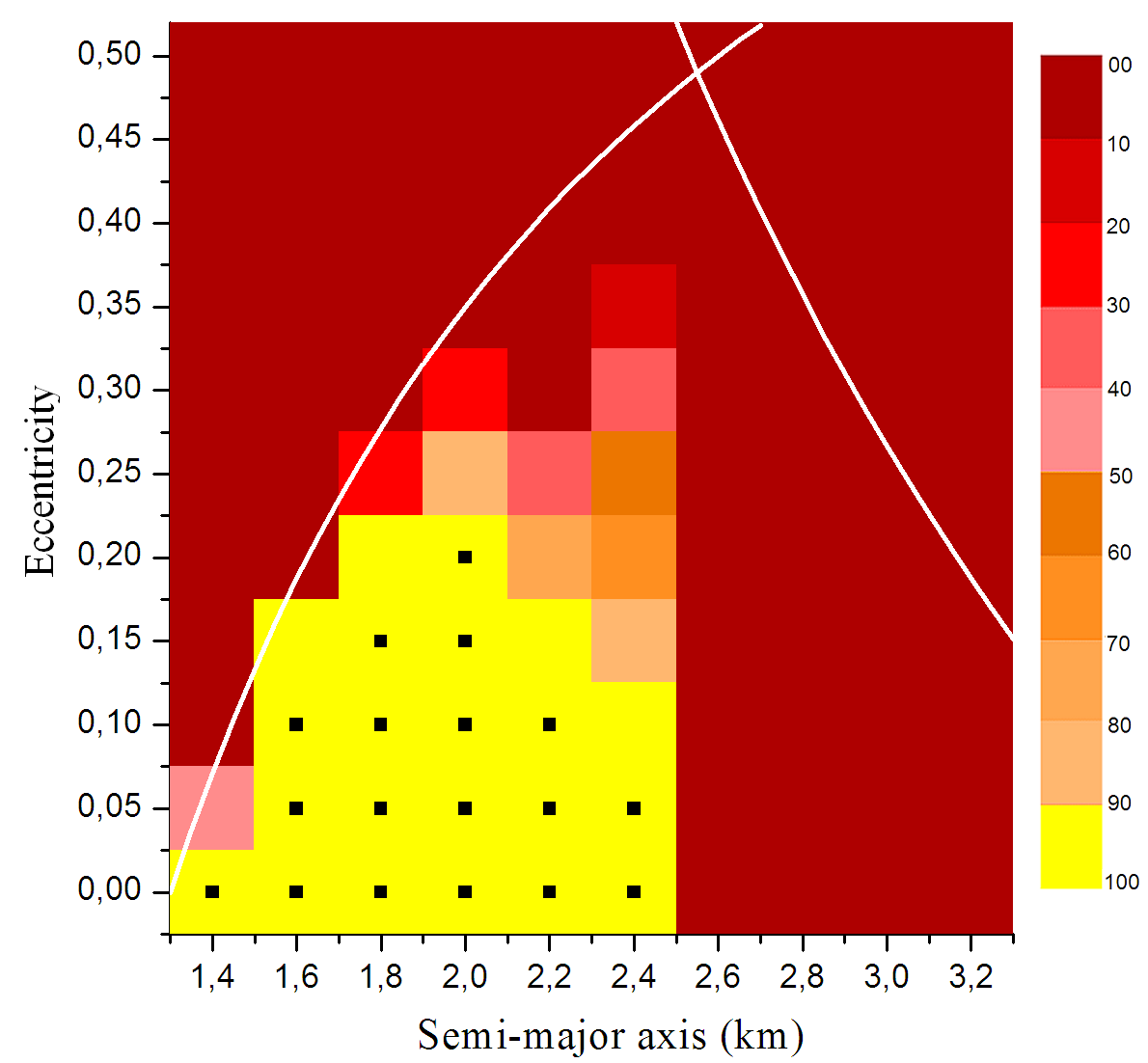}}\qquad
\hspace{0.0cm}
\vspace{0.6cm}
\subfigure[]{\includegraphics[height=7.5cm]{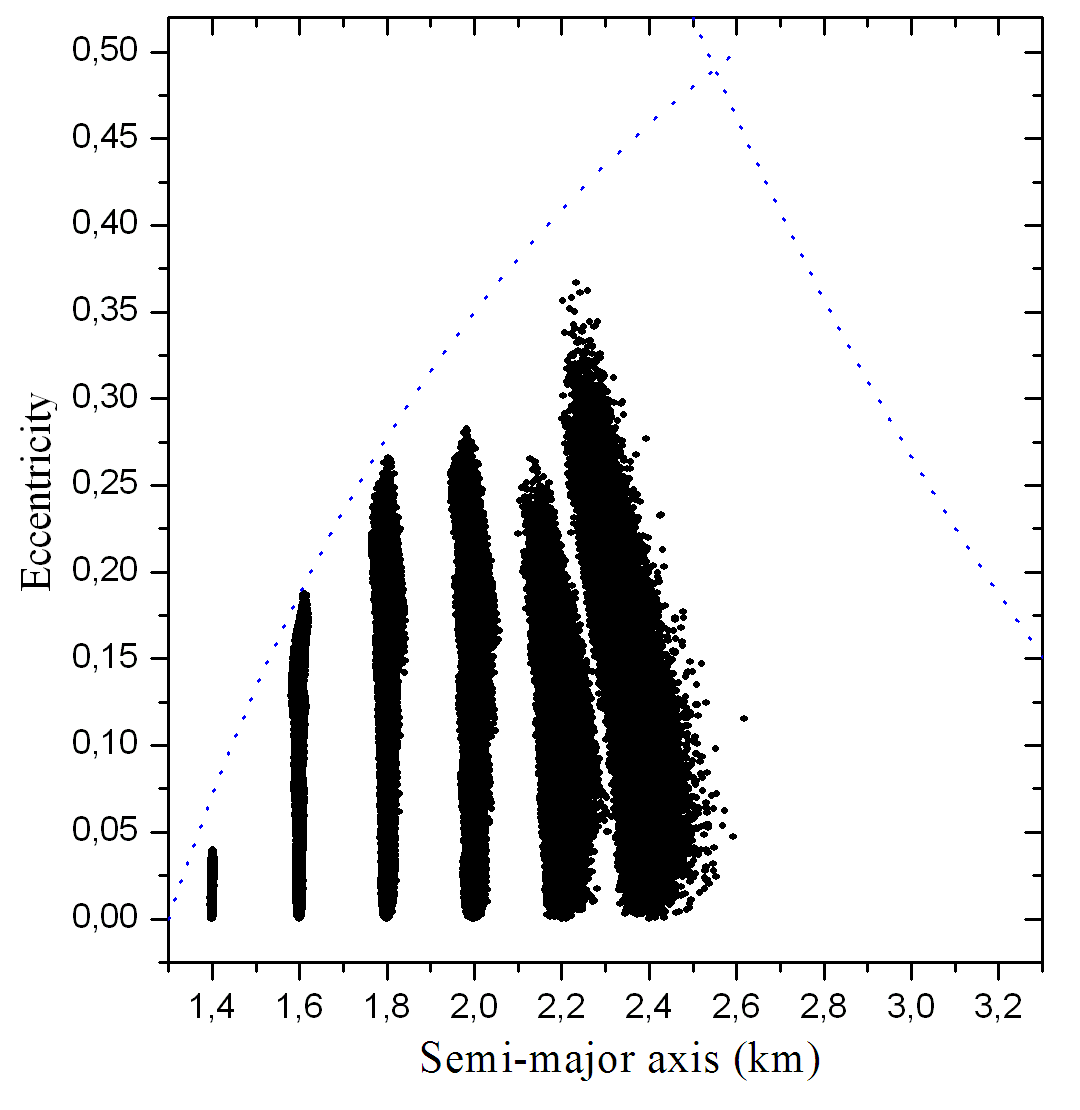}}}
\caption{a) Diagram of stability of region 1, for a time span of 2 years. The scale goes from $0.0\%-10.0\%$ of the particles that survive in that 
region (dark red) to $90\%-100\%$ of survivors (yellow). The yellow boxes marked with the 
small black point indicate the cases of $100\% of survival$. The white lines indicates the limits of the region. On the left is the collision-line with Alpha and 
on the right is the collision-line with Gamma, denoting the ejection distance $d$. b) Scattering of particles in region 1. It shows the evolution of $a$ and $e$, 
along the 2 years of integration, for every output step, for the particles belonging to the region where there were $90\%-100\%$ of survival.
The blue lines indicates the limits of the region.}
\label{fig_re1}
\end{figure*}

In region 1, the particles are orbiting Alpha with the orbital elements: $1.4\leq a \leq 3.2$ (km) taken every $0.2$ km, $0.0 \leq e \leq 0.50$ taken every $0.05$,  
and $100$ particles for each pair $(a\times e $), with random values for $f, \omega, \Omega$, as described in section \ref{sec_initial}. Such combination of values 
resulted in a total of $11,000$ particles placed in region 1, with $I=0,0^{\circ}$.

The diagram of Figure \ref{fig_re1}a shows the result found. It was considered a grid of semi-major axis versus eccentricity, and each one of the small ``boxes''
holds the information of $100$ particles that share the same initial values for $a$ and $e$. 
 
On such diagram it is shown how much particles survived for each set of initial conditions. The coded color indicates the percentage of survivors. 
The color yellow indicates the initial conditions for what $90\%-100\%$ of particles survive for $2$ years.
The yellow boxes marked with the small black point indicate the specific cases where $100\%$ of the particles survive (stability).
The dark red color indicates the case where less than  $10\%$ of the particles survived for the same period.

As discussed in subsection \ref{sec_method}, if a particle in region 1 exceeds the distance of $3.804$ km from Alpha, it is considered ejected. On the other hand,
if such distance is smaller than the radius of Alpha, we have collision. So, we define the limits of the region, represented on the diagram of Figure \ref{fig_re1}a by the 
white lines. On the right is the collision-line with Alpha. It gives the limit from which the pericentre $(q)$ of the orbit of the particle is smaller than the radius of Alpha 
($R_{Alpha}$). Being $q=a(1-e)$, then, from the collision conditions we have that $q \leq R_{Alpha}$, leading to the relation: $a\leq R_{Alpha}/(1-e)$, with $0.0\leq e \leq 0.5$.
On the left is the collision-line with Gamma, denoting the limit from which the apocenter $(Q)$ of the orbit of the particle crosses the orbit of Gamma, 
and by the definition, is beyond of the ejection distance $d$. 
Being $Q=a(1+e)$, then, the relation $a \geq d/(1+e)$, with $0.0\leq e \leq 0.5$, gives the limit of the collision-line with Gamma, for particles in region 1.

We see from the diagram of Figure \ref{fig_re1}a that in region 1, the stability is found on the region closer to Alpha, for lower values of eccentricity. 
As the value of semi-major axis increases, the collisions with Gamma, or the ejections, become more frequent, given rise to instability.  
Similar behavior happens when the value of eccentricity increases and the orbit of the particles approach the collision-line with Alpha.

The diagram of Figure \ref{fig_re1}b shows the scattering of the particles in region 1. It shows the evolution of $a$ and $e$ 
along the 2 years of integration, for every output step, for the surviving particles
belonging to the region where there were $90\%-100\%$ of survival (yellow). 
We see that, in fact, the particles are confined within the limits
of the region (blue lines). Besides that, there is almost no variation of the semi-major axis. The external particles are the most perturbed, as expected.

The Figure \ref{fig_estatistics}a shows the percentage of collisions, ejections and survivors, for the internal and external regions.
The Figure \ref{fig_estatistics}b shows the percentage of collisions of the particles with each one of the bodies of the triple system in such regions. 
From these figures we see that the ejections of particles are not so significant in region 1. Due to their proximity with both massive asteroids,
Alpha and Gamma, the collisions are more frequent, mainly with Alpha.


\subsection{Region 2}
\label{subsec_stabregion2}

\begin{figure*}
\subfigure[]{\includegraphics[scale=0.11]{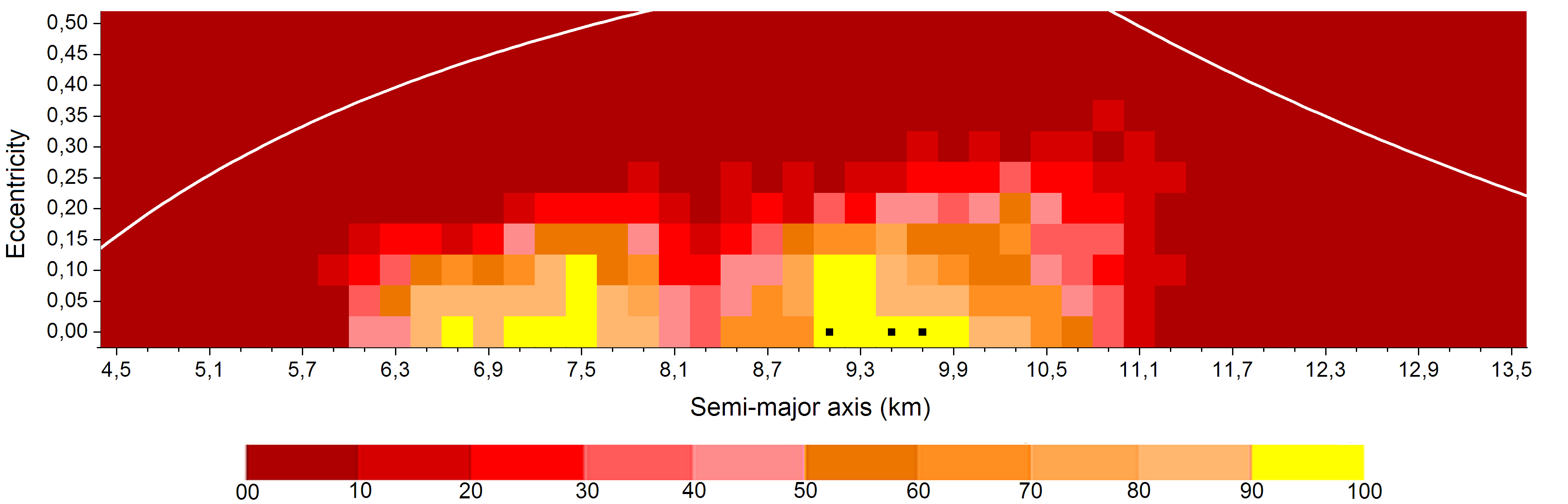}}
\hspace{-0.2cm}
\subfigure[]{\includegraphics[scale=0.11]{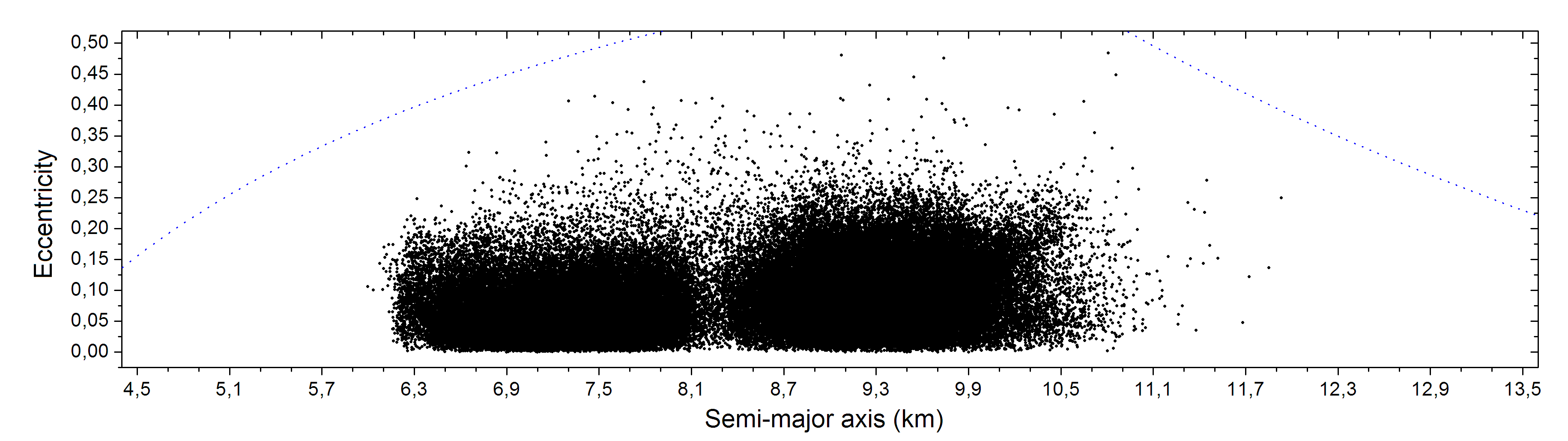}}
\caption{a) Diagram of stability of region $2$, for a time span of 2 years. The scale goes from $0.0\%-10.0\%$ of the particles that survive in that 
region (dark red) to $90\%-100\%$ of survivors (yellow).The yellow boxes marked with the small black point indicate the cases of $100\%$ of survival.
The white and blue lines indicate the limits of the region. On the left is the collision-line with Gamma, for particles in region 2. 
On the right is the collision-line with Beta, denoting the ejection distance. b) Scattering of the particles in region 2.
It shows the evolution of $a$ and $e$,
along the 2 years of integration, for every output step, for the particles belonging to the region where there were $90\%-100\%$ of survival.
 The blue lines indicates the limits of the region.}
\label{fig_region2}
\end{figure*}

In region 2, the particles are orbiting the asteroid Alpha with semi-major axis $4.5 \leq a \leq 13.5$ (km) taken every $0.2$ km, and $0.0 \leq e \leq 0.50$ taken every $0.05$.
For each pair $(a\times e)$, there were $100$ particles with random values for $f, \omega, \Omega$, with initial inclination  $I=0,0^{\circ}$. 
Such combination of values resulted in a total of $50,600$ particles placed in region 2. 

As the previous diagram of Figure \ref{fig_re1}, the diagram of Figure \ref{fig_region2} shows the region of stability found for region $2$. As before, a grid of
semi-major axis versus eccentricity was adopted. The details about such grid are in subsection \ref{subsec_stabregion1}.

As discussed in subsection \ref{sec_method}, if a particle in region 2 exceeds the distance of $d=16.633$ km from Alpha, it is considered ejected. On the other hand,
for distance near $3.804$ km, the particles cross the orbit of Gamma, increasing the collision probability, even with Alpha.
We represent such limits by the white lines on the diagram of Figure \ref{fig_region2}a. On the left is the collision-line with Gamma, for particles in region 2. 
It gives the limit from which the pericentre $(q)$ of the orbit of the particle is smaller than $d=3.804$ km.
Being $q=a(1-e)$, then we have the relation: $a\leq d/(1-e)$, with $0.0\leq e \leq 0.5$.
On the right is the collision-line with Beta, denoting the limit from which the apocenter $(Q)$ of the orbit of the particle crosses the orbit of Beta, 
and by the definition, is beyond of the ejection distance $d=16.633$ km.
Being $Q=a(1+e)$, then, the relation $a \geq d/(1+e)$, with $0.0\leq e \leq 0.5$, gives the limit of the collision-line with Beta, for particles in region 2.

We see from the diagram of Figure \ref{fig_region2}a that almost no particle survives 2 years in the region between the asteroids Gamma and Beta. The better results are found
approximately in the middle of the region for the lower values of eccentricities, and the condition of $100\%$ of survival
was reached only for three specific values of semi-major axis ($a=9.1, 9.5, 9.7$ km), for $e=0.0$. The observed instability can be explained by the fact that region 2
is a region surrounded by massive bodies, and also by the presence of resonant motion of the particles with Gamma or Beta, characterized by the gap near the center 
of the region. This is discussed in section \ref{sec_res}.

The diagram of Figure \ref{fig_region2}b shows the scattering of the particles in region 2. It shows the evolution of $a$ and $e$ of the surviving particles,
belonging to the region where there were $90\%-100\%$ of survival (yellow), along the 2 years of integration, for every output step. Again we confirm that the particles are confined within the limits
of the region (blue lines). 

In region 2 the collisions still prevail over the ejections, as can be seen in Figure \ref{fig_estatistics}a, and the collisions with Beta are more frequent	
(Figure \ref{fig_estatistics}b).

\subsection{Region 3}
\label{subsec_stabregion3}

\begin{figure*}
\mbox{%
\subfigure[]{\includegraphics[height=6.5cm]{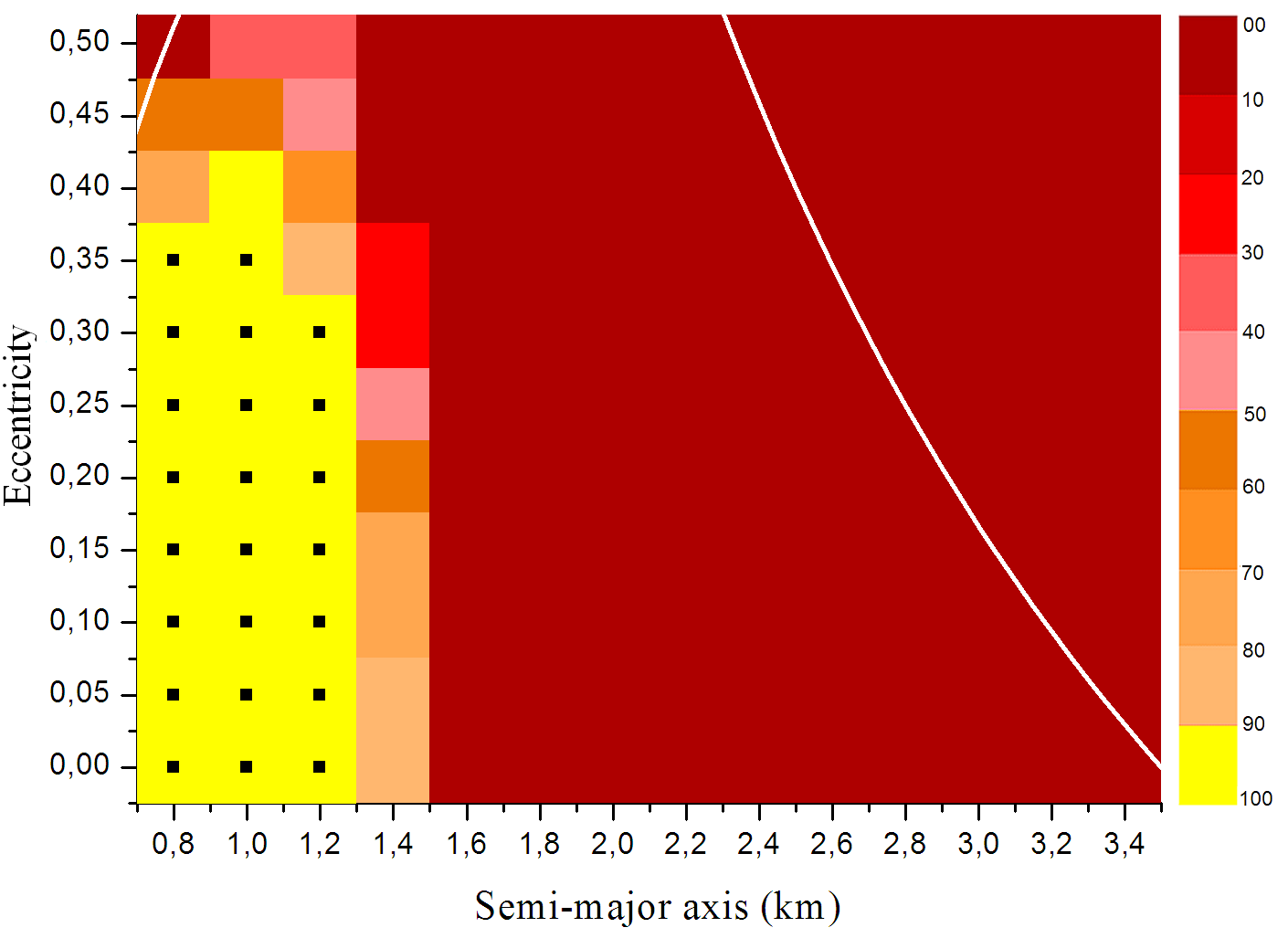}}\qquad
\hspace{-0.5cm}
\subfigure[]{\includegraphics[height=6.5cm]{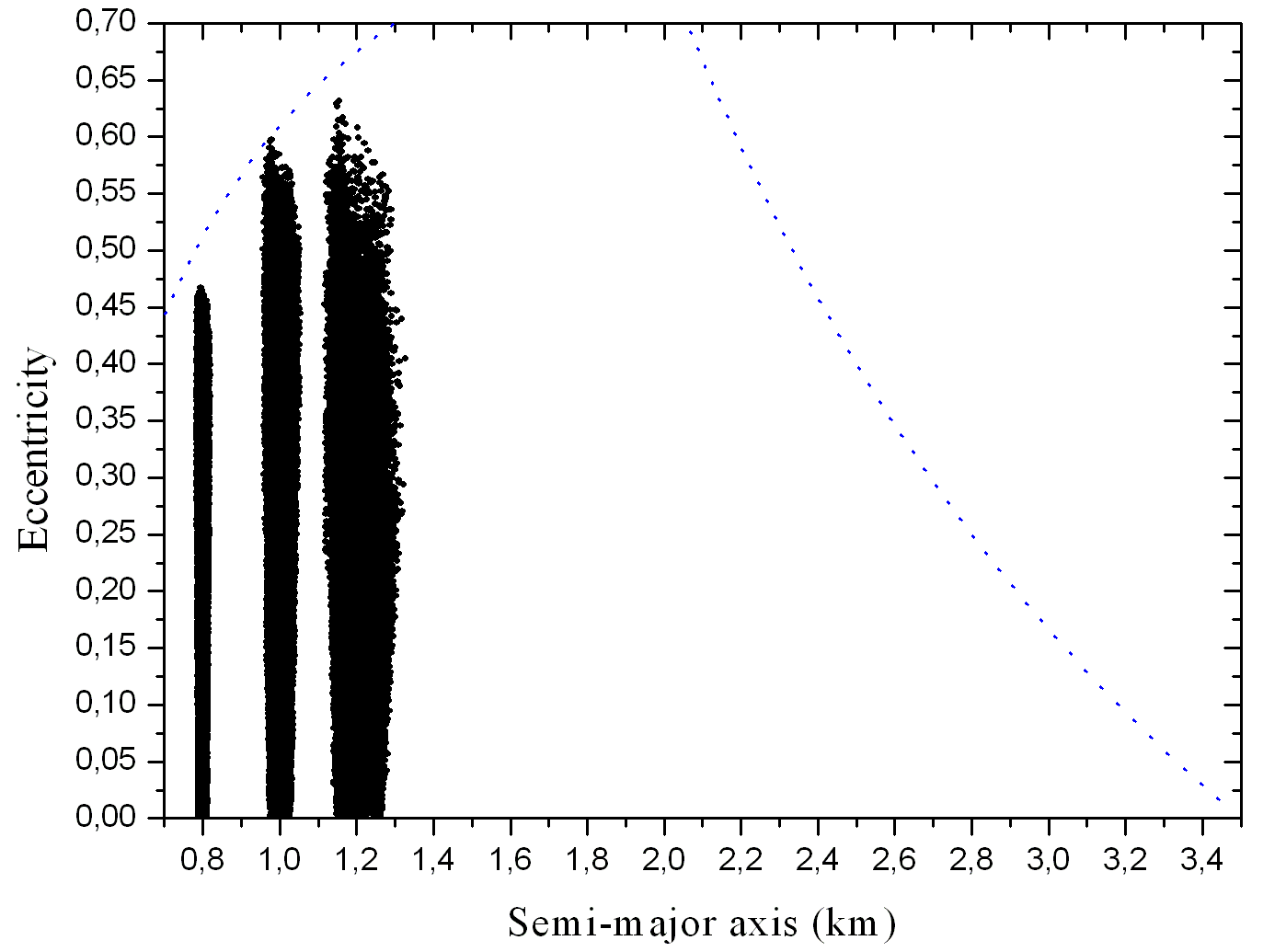}}}
\caption{a) Diagram of stability of region $3$, for a time span of 2 years. The scale goes from $0.0\%-10.0\%$ of the particles that survive in that 
region (dark red) to $90\%-100\%$ of survivors (yellow).
The yellow boxes marked with the small black point indicate the cases of 100\% of survival.
The white lines indicates the limits of the region. On the left is the collision-line with Beta, and on the right is the ejection-line.
b) Scattering of particles in region 3. It shows the evolution of $a$ and $e$,
along the 2 years of integration, for every output step, for the particles belonging to the region where there were $90\%-100\%$ of survival.
 The blue lines indicates the limits of the region.}
\label{fig_region3}
\end{figure*}

In region 3, the particles are orbiting the asteroid Beta with semi-major axis $0.8 \leq a \leq 3.4$ (km) taken every $0.2$ km, and $0.0 \leq e \leq 0.50$ taken every $0.05$.
For each pair $(a\times e)$, there were $100$ particles with random values for $f, \omega, \Omega$, with initial inclination  $I=0,0^{\circ}$. 
Such combination of values resulted in a total of $15,400$ particles placed in region 3. 

The diagram of Figure \ref{fig_region3}a shows the region of stability found for region $3$. As before, a grid of semi-major axis versus eccentricity was adopted. 
The details about such grid are in subsection \ref{subsec_stabregion1}. Similar to what happens in region 1, the region where more particles survive in region 3 is that closer to 
the asteroid that they orbit, in this case, the asteroid Beta. 

As discussed in subsection \ref{sec_method}, if a particle on region 3 exceeds the distance of $3.4$ km from Beta, it is considered ejected. On the other hand,
if such distance is smaller than the radius of Beta, we have collision. So, we define the limits of the region, represented on the diagram of Figure \ref{fig_re1}b by the 
white lines. On the right is the collision-line with Beta. It gives the limit from which the pericentre $(q)$ of the orbit of the particle is smaller than the radius of Beta
($R_{Beta}$). Being $q=a(1-e)$, then, from the collision conditions we have that $q \leq R_{Beta}$, leading to the relation: $a\leq R_{Beta}/(1-e)$, with $0.0\leq e \leq 0.5$.
On the left is the ejection-line, denoting the limit from which the apocenter $(Q)$ of the orbit of the particle is greater than the ejection distance $d=3.4$ km.
Being $Q=a(1+e)$, then, the relation $a \geq d/(1+e)$, with $0.0\leq e \leq 0.5$, gives the limit of ejection for particles in region 3.

As presented for regions 1 and 2, the diagram of Figure \ref{fig_region3}b shows the scattering of the particles in region 3 along the 2 years of 
integration, for every output step.  
We see that there is almost no variation of the semi-major axis. The external particles are the most perturbed, and almost no particles survive in the region 
beyond half Hill's radius ($\approx1.7$ km).

\begin{figure}
\subfigure[]{\includegraphics[scale=0.24]{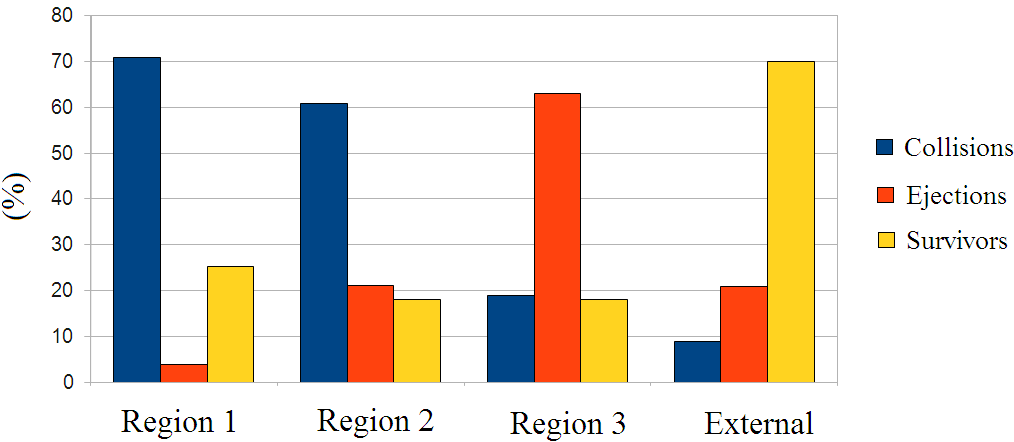}}
\subfigure[]{\includegraphics[scale=0.25]{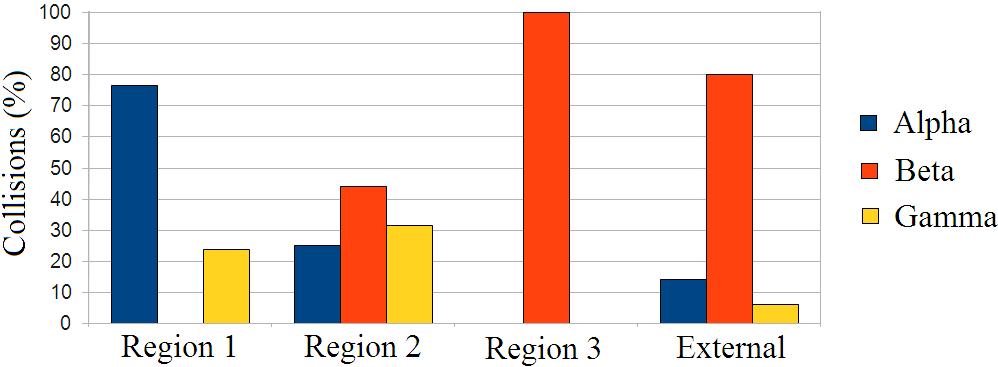}}
\caption{a)Percentage of collisions, ejections and of remaining particles for each region. b) Percentage of collisions with each component of the triple system.} 
\label{fig_estatistics}
\end{figure}

According to Figure \ref{fig_estatistics}a, the number of ejections increases in region 3, which makes sense, since Beta is farthest 
body of the system. Due to the definition of the region 3, the collisions may happen only with Beta (Figure \ref{fig_estatistics}b).

\section{Inclined Prograde Cases}

Now that we have discussed the planar case, we are going to consider inclined particles orbiting the same previous internal regions. The purpose is analyze what change
on those stable regions found in the planar case (section \ref{sec_results}) when the particles have an inclination relative to the equator of the central body.
For all the three internal regions, the initial conditions and the number of particles are exactly the same, except by the inclination of the particles
that was taken from  $15.0^{\circ}$ until $90.0^{\circ}$, in a fixed interval of $15.0^{\circ}$ (six cases for each region). The results found are presented in the next subsections.

\begin{figure*}
\mbox{%
\subfigure[]{\includegraphics[height=6.5cm]{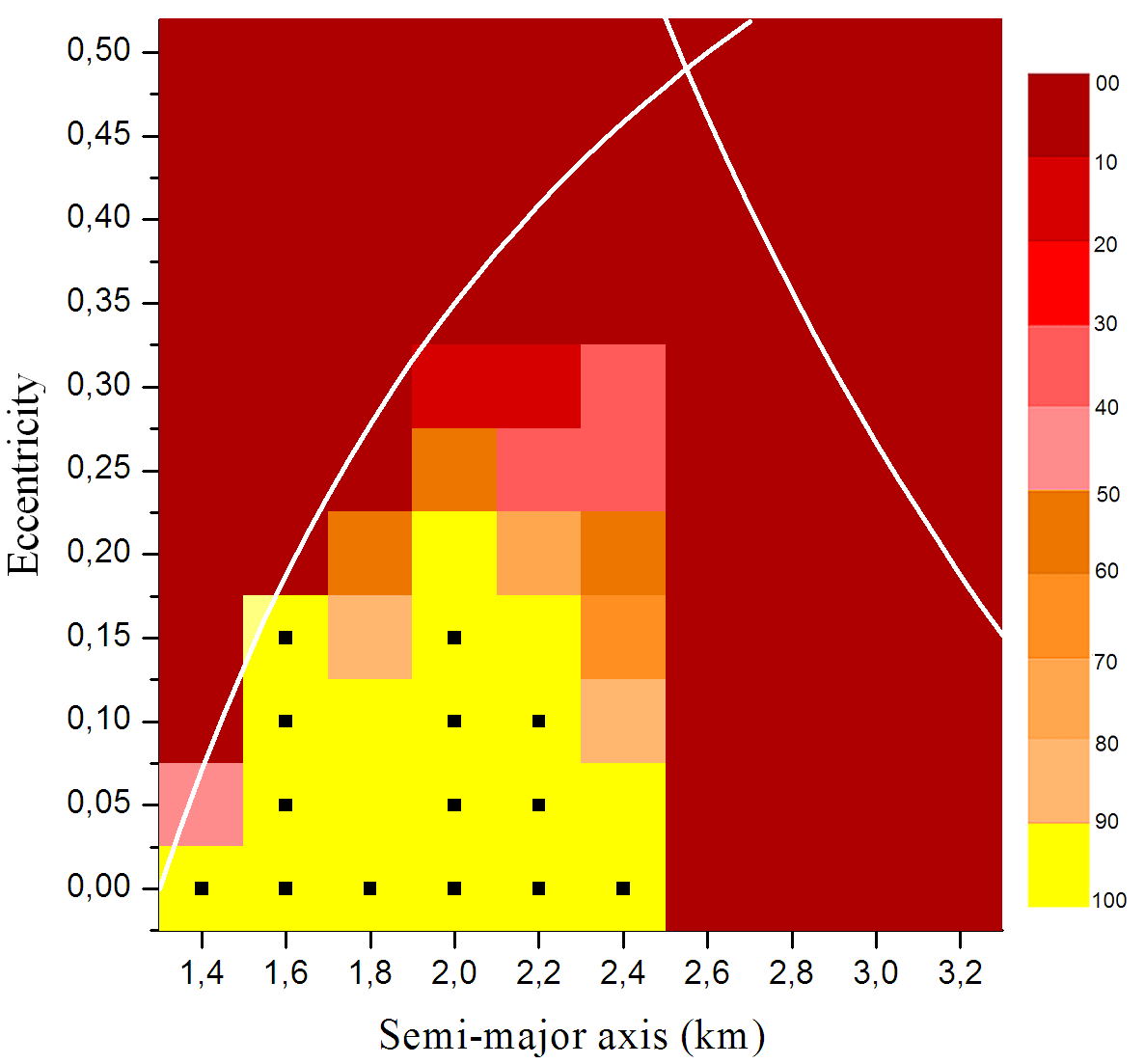}}\qquad
\subfigure[]{\includegraphics[height=6.5cm]{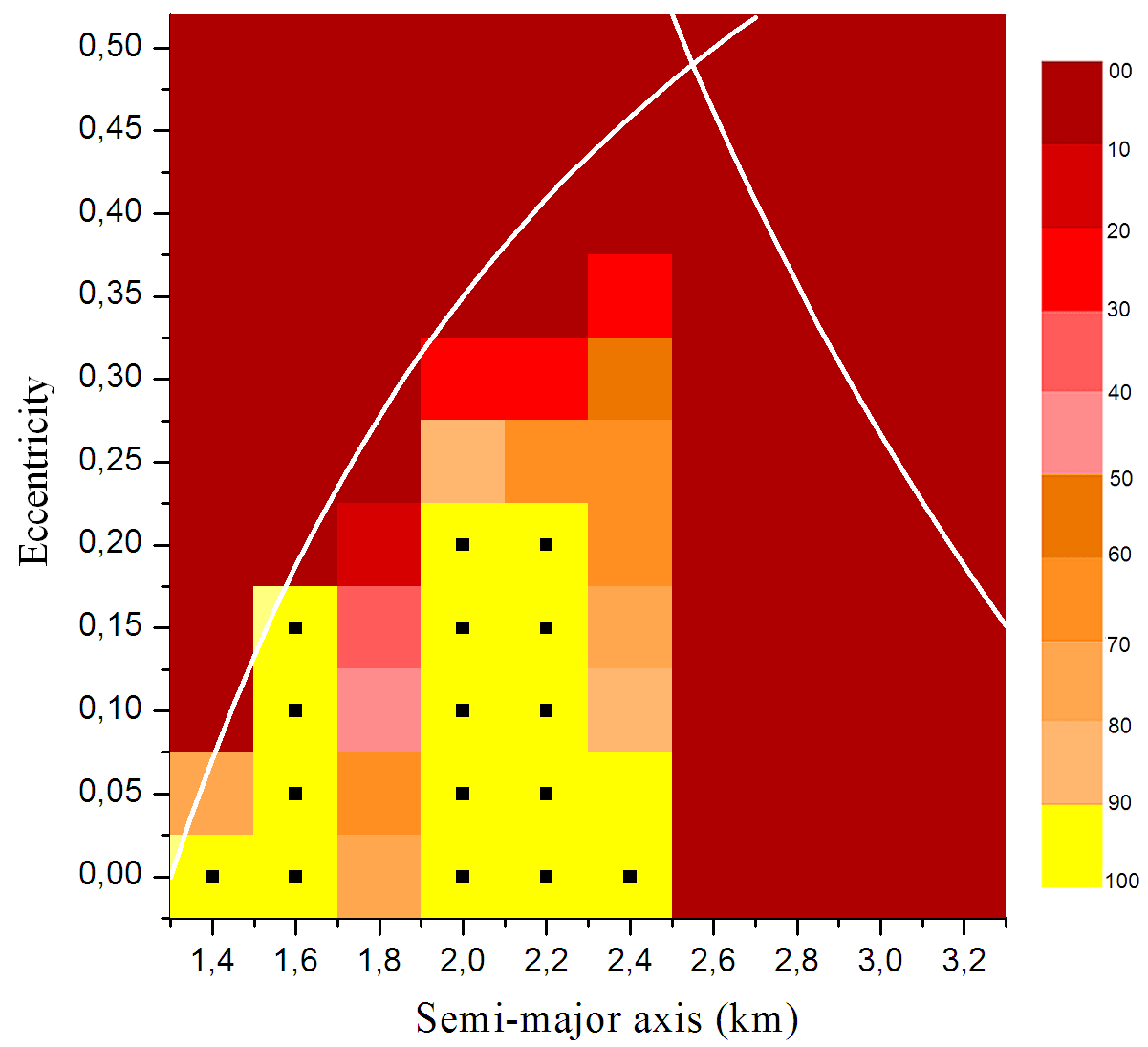}}}
\mbox{%
\subfigure[]{\includegraphics[height=6.5cm]{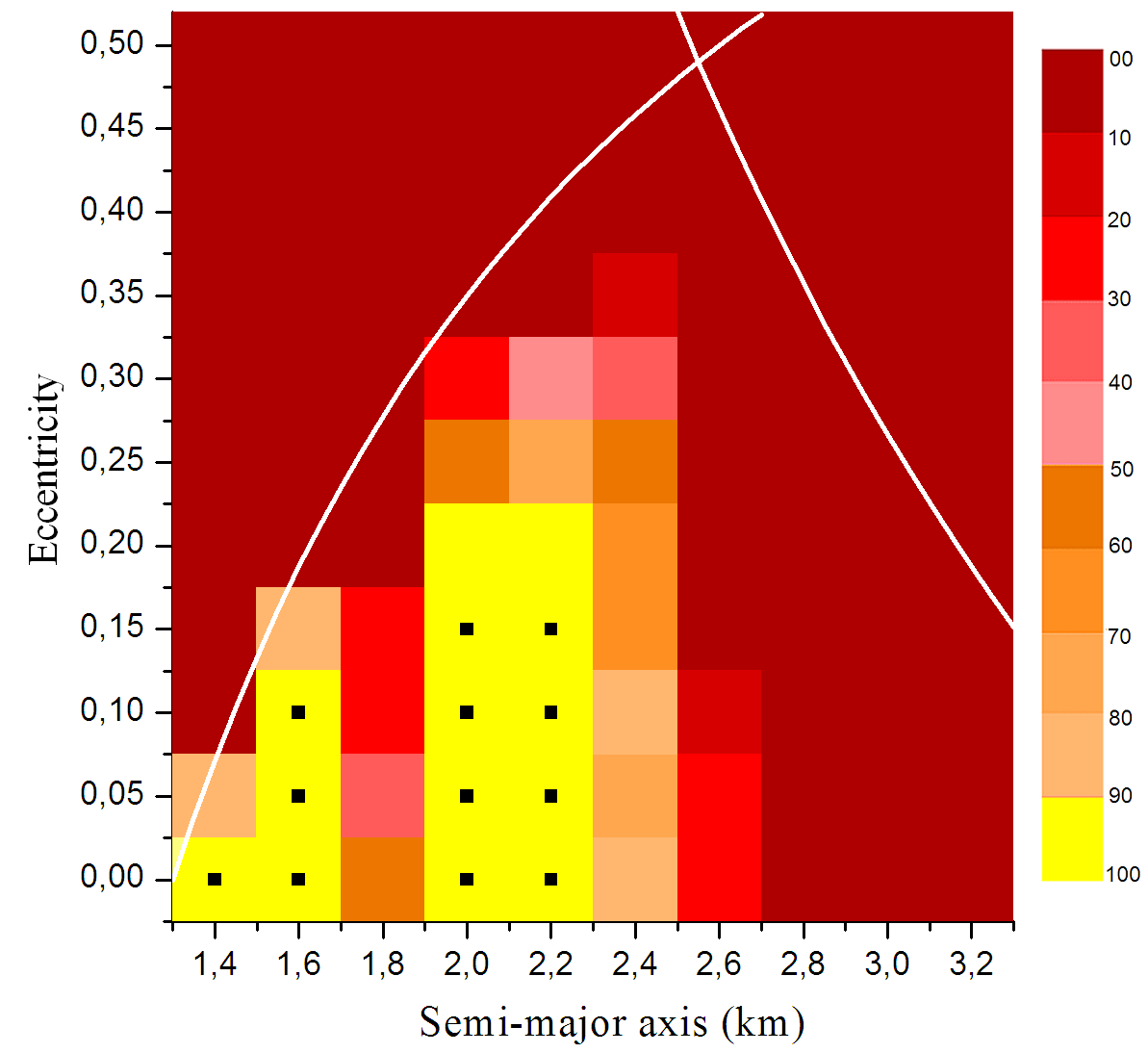}}\qquad
\subfigure[]{\includegraphics[height=6.5cm]{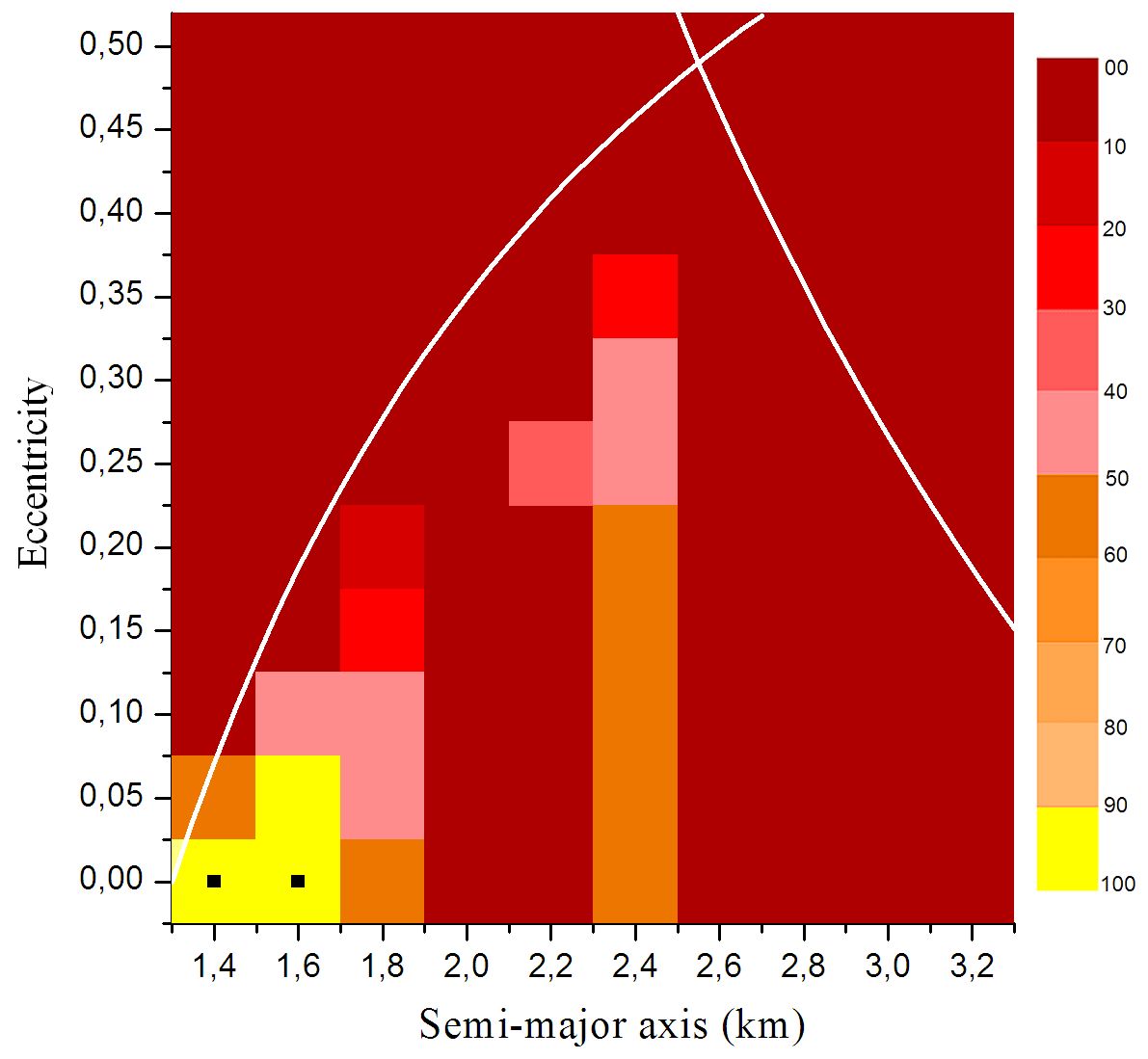}}}
\mbox{%
\subfigure[]{\includegraphics[height=6.5cm]{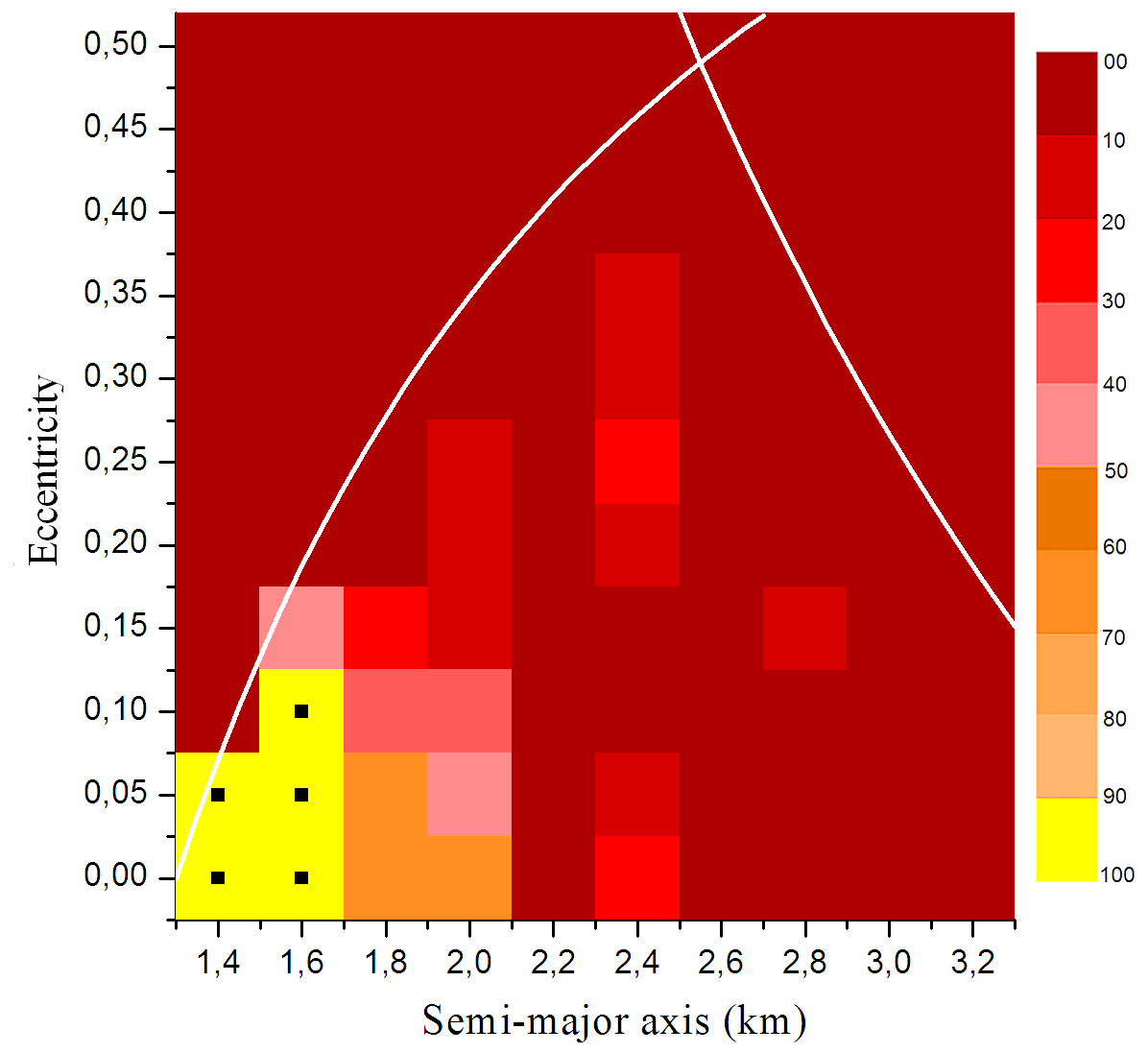}}\qquad
\subfigure[]{\includegraphics[height=6.5cm]{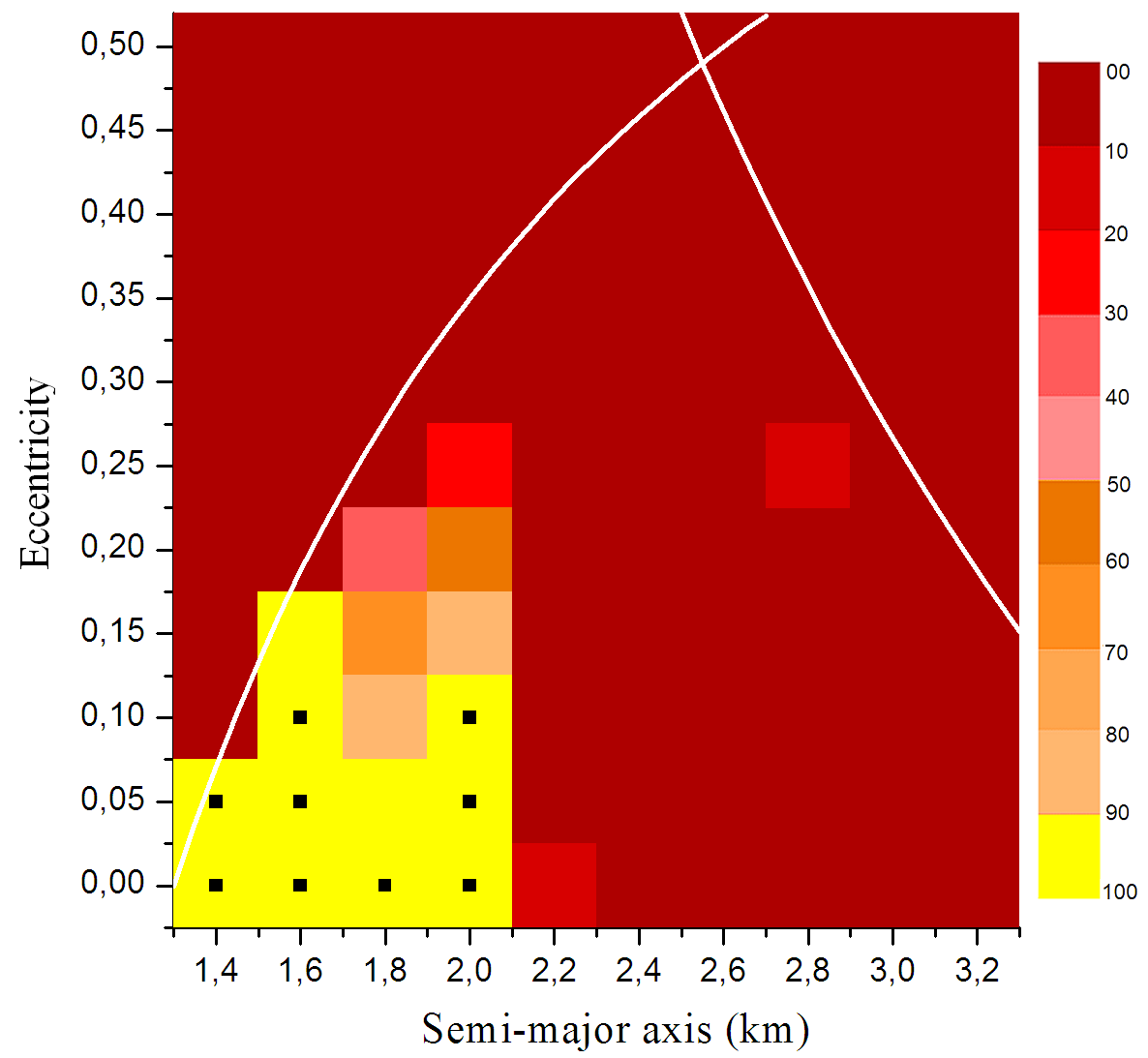}}}
\caption{Diagram of stability of region $1$, for a time span of 2 years. a) $I=15.0^{\circ}$, b) $I=30.0^{\circ}$, c) $I=45.0^{\circ}$, 
d) $I=60.0^{\circ}$, e) $I=75.0^{\circ}$, f) $I=90.0^{\circ}$. The white lines indicate the limits of the region.The yellow boxes marked with the small
 black point indicate the cases of 100\% of survival.}
\label{fig_incl_re1}
\end{figure*}

\begin{figure*}
\subfigure{\includegraphics[height=3.48cm]{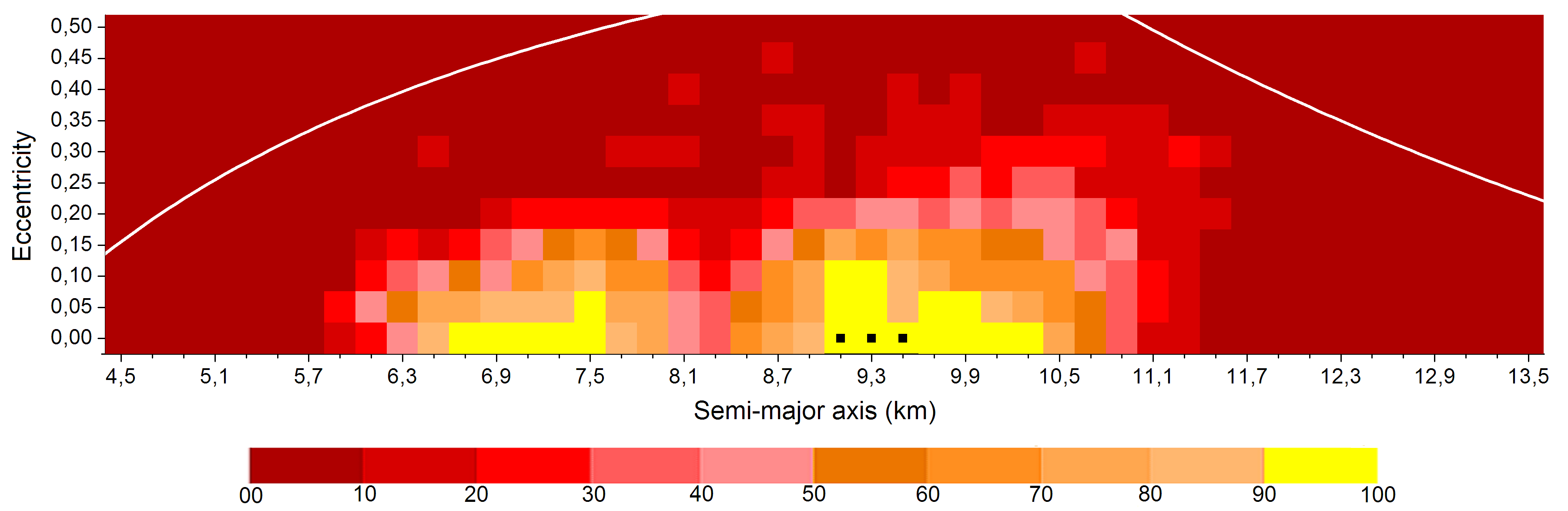}}
\subfigure{\includegraphics[height=3.5cm]{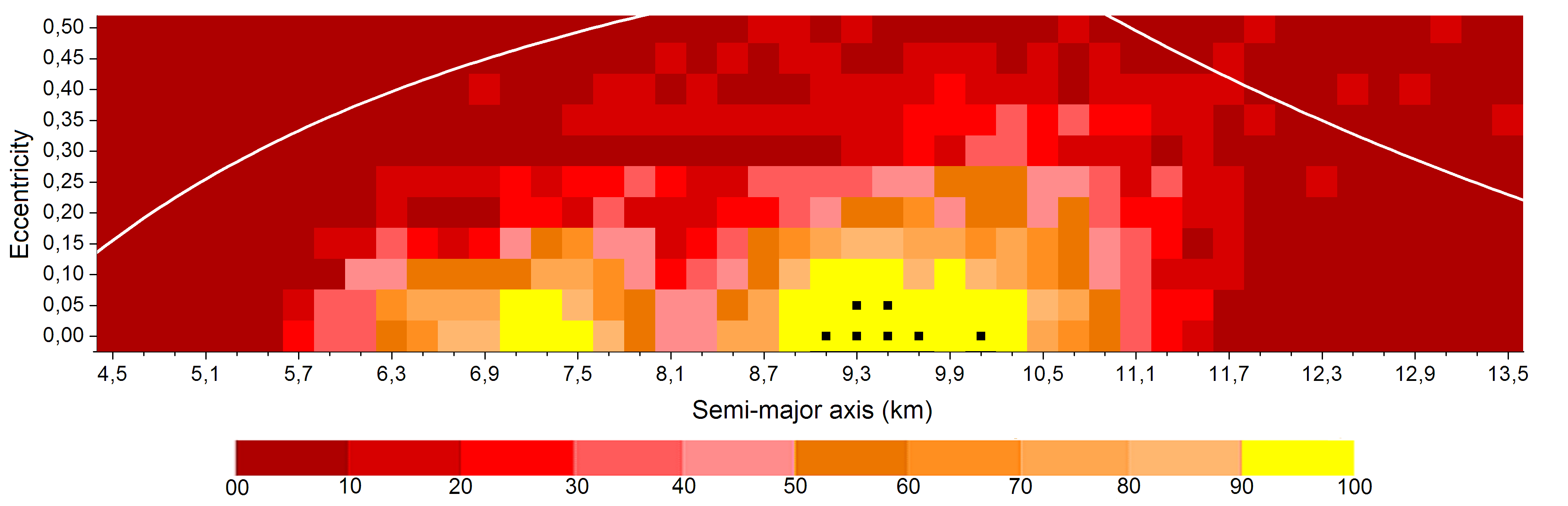}}
\subfigure{\includegraphics[height=3.35cm]{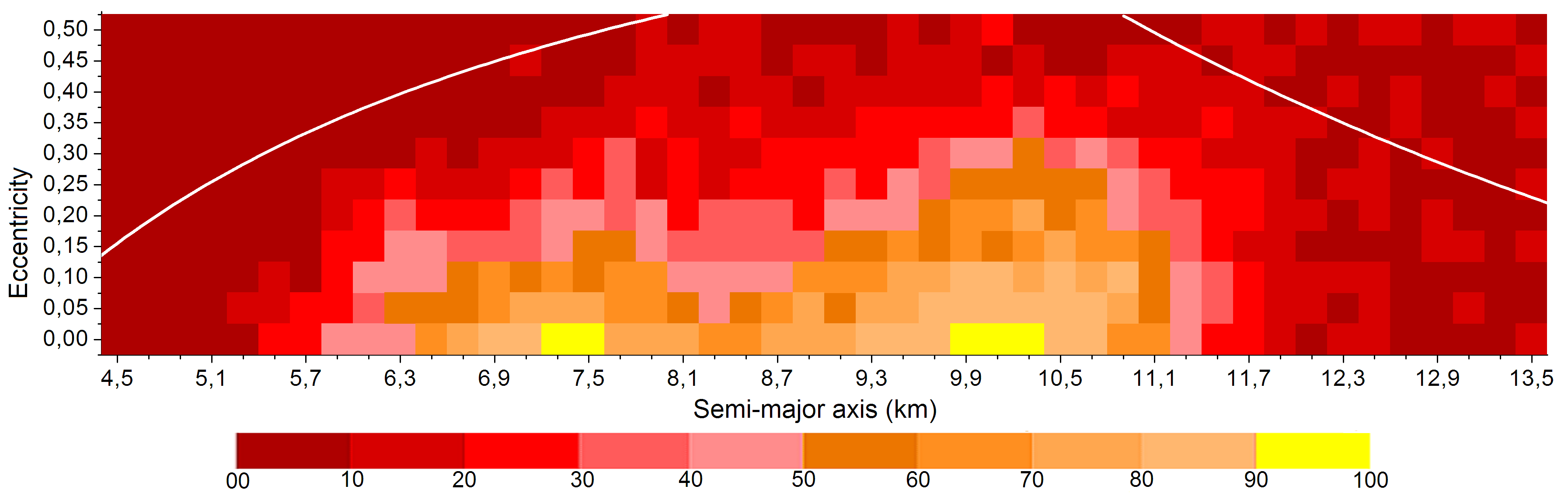}}
\subfigure{\includegraphics[height=3.5cm]{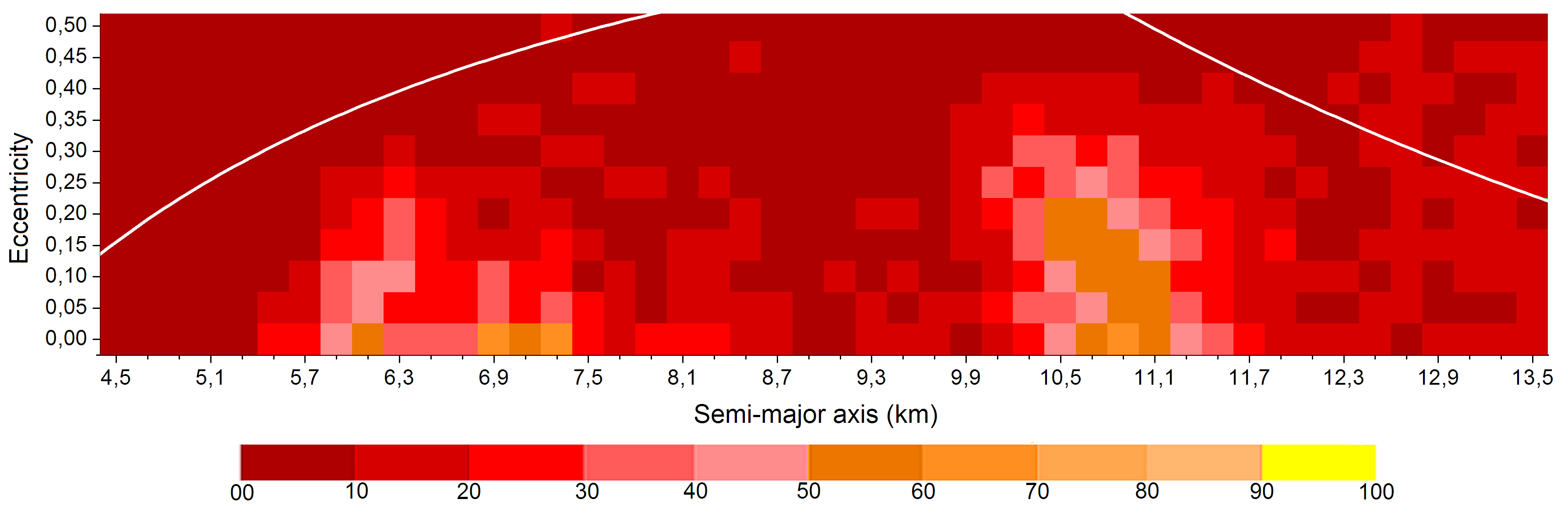}}
\subfigure{\includegraphics[height=3.5cm]{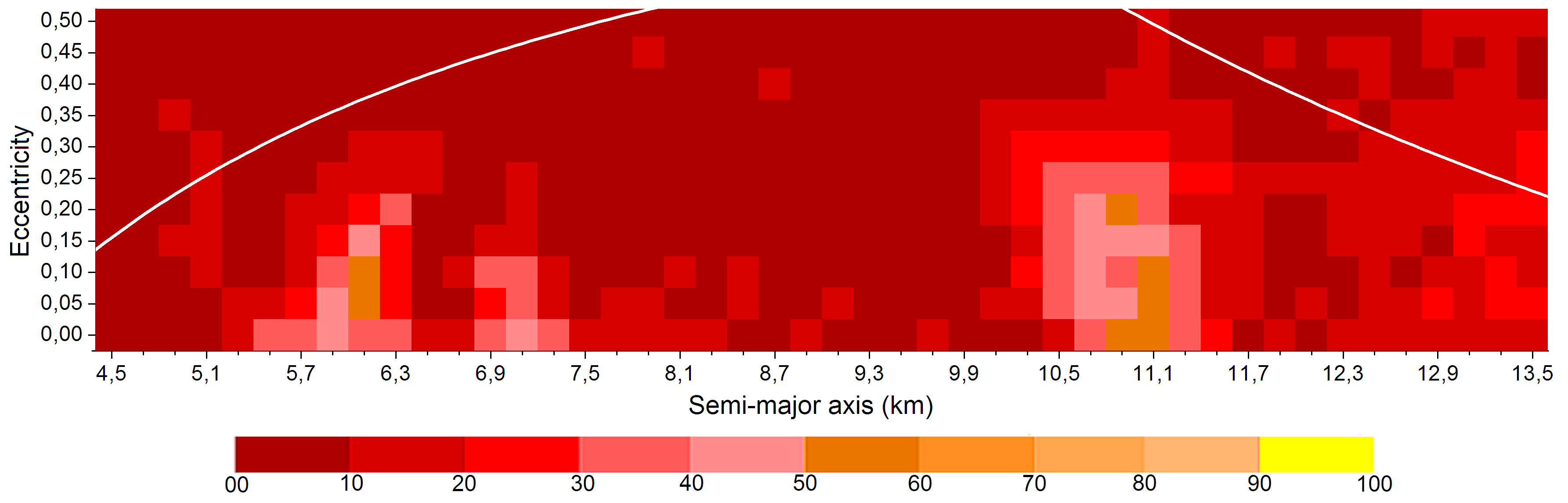}}
\subfigure{\includegraphics[height=3.6cm]{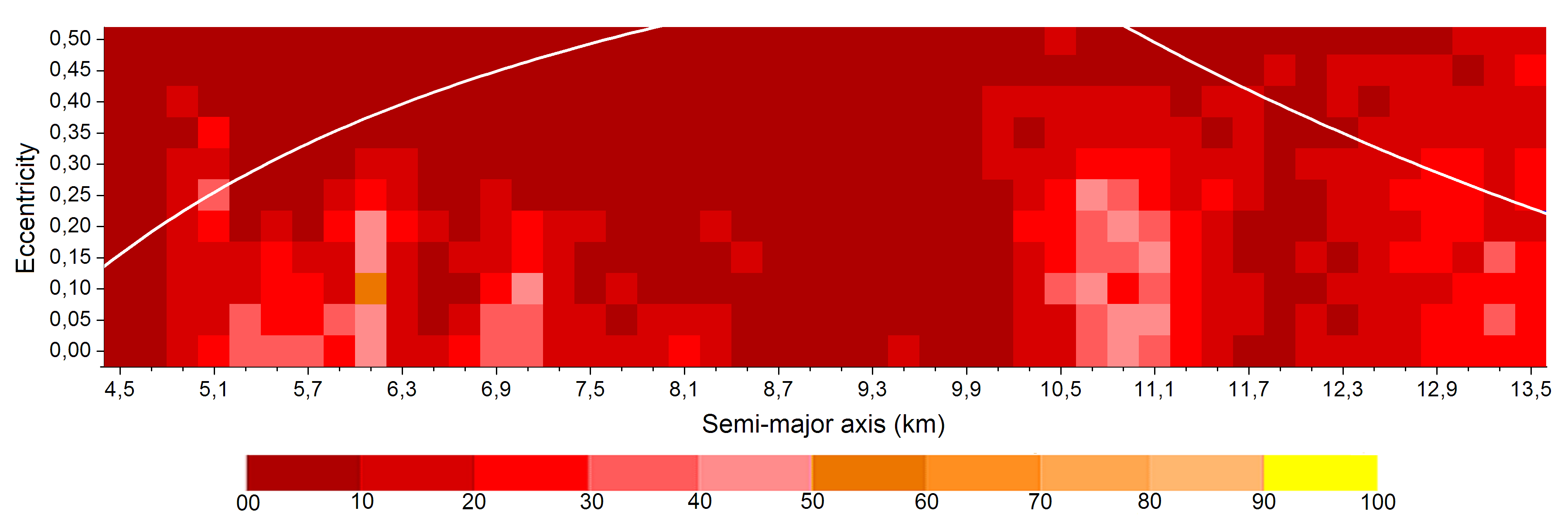}}
\caption{Diagram of stability of region $2$, for a time span of 2 years for $I=15.0^{\circ}, 30.0^{\circ},I=45.0^{\circ}, I=60.0^{\circ}, 
I=75.0^{\circ}$ and $I=90.0^{\circ}$. The white lines indicate the limits of the region.The yellow boxes marked with the small
 black point indicate the cases of 100\% of survival.}
\label{fig_incl_re2}
\end{figure*}

\begin{figure*}
\mbox{%
\subfigure[]{\includegraphics[height=5.2cm]{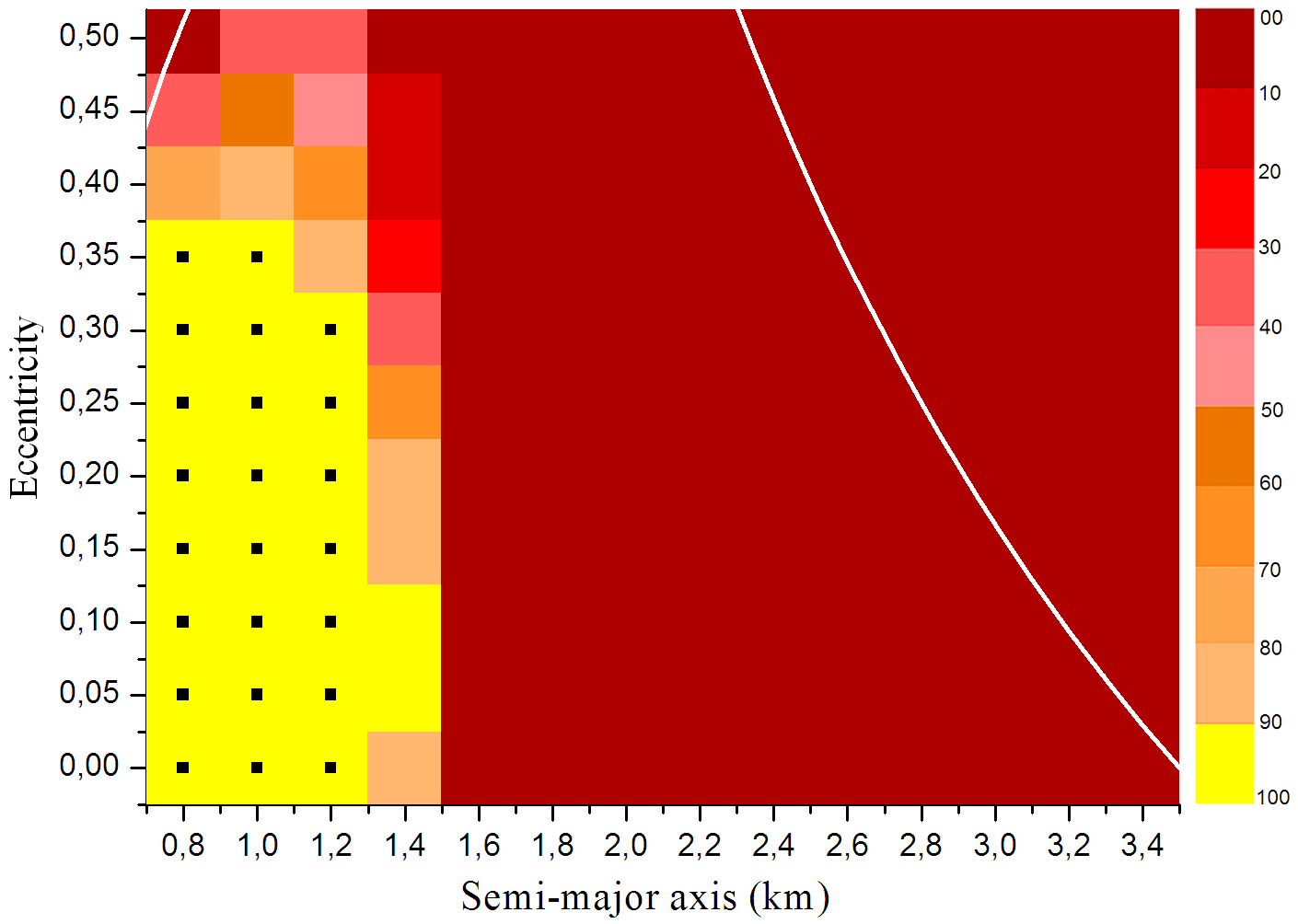}}\qquad
\subfigure[]{\includegraphics[height=5.2cm]{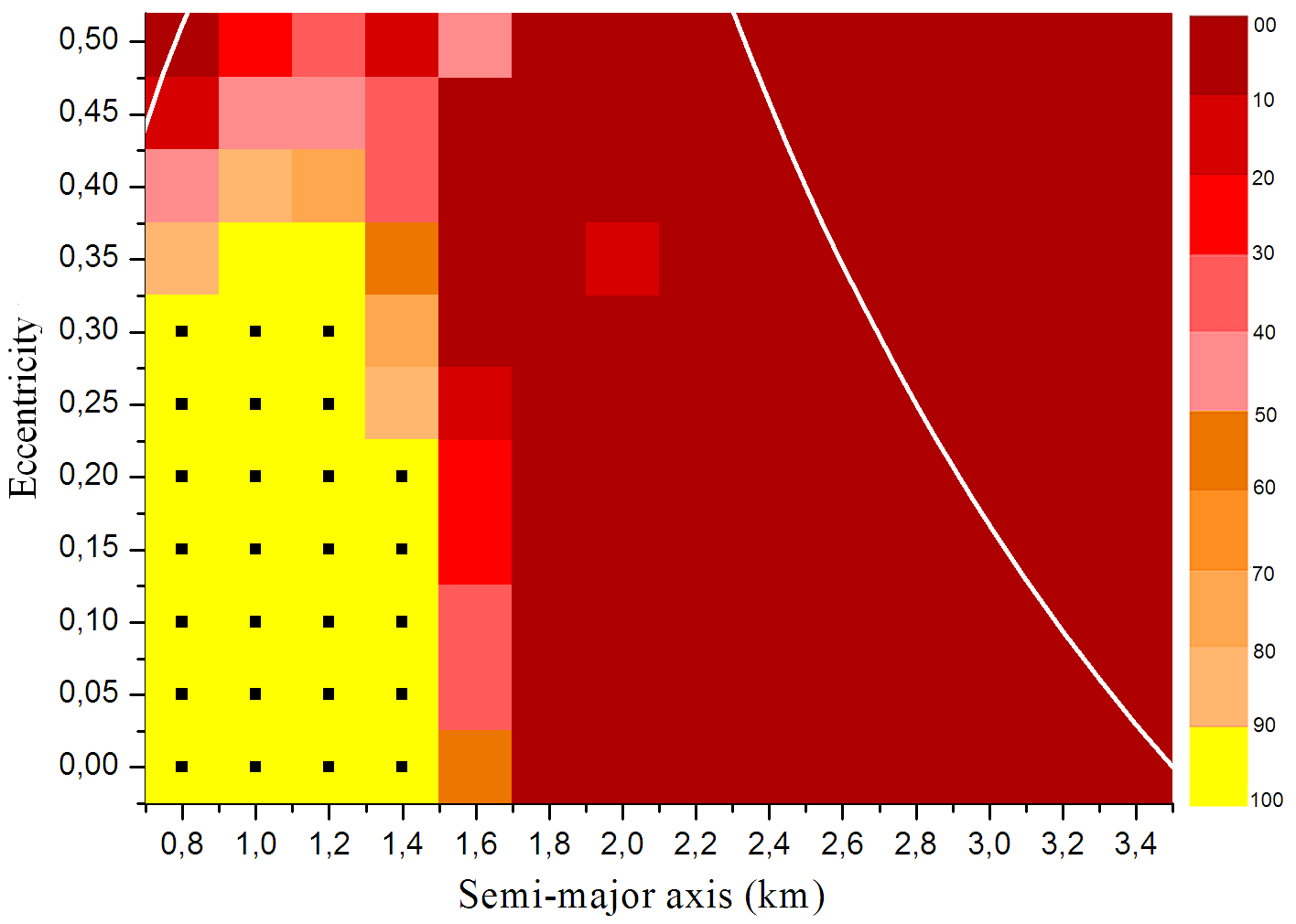}}}
\mbox{%
\subfigure[]{\includegraphics[height=5.2cm]{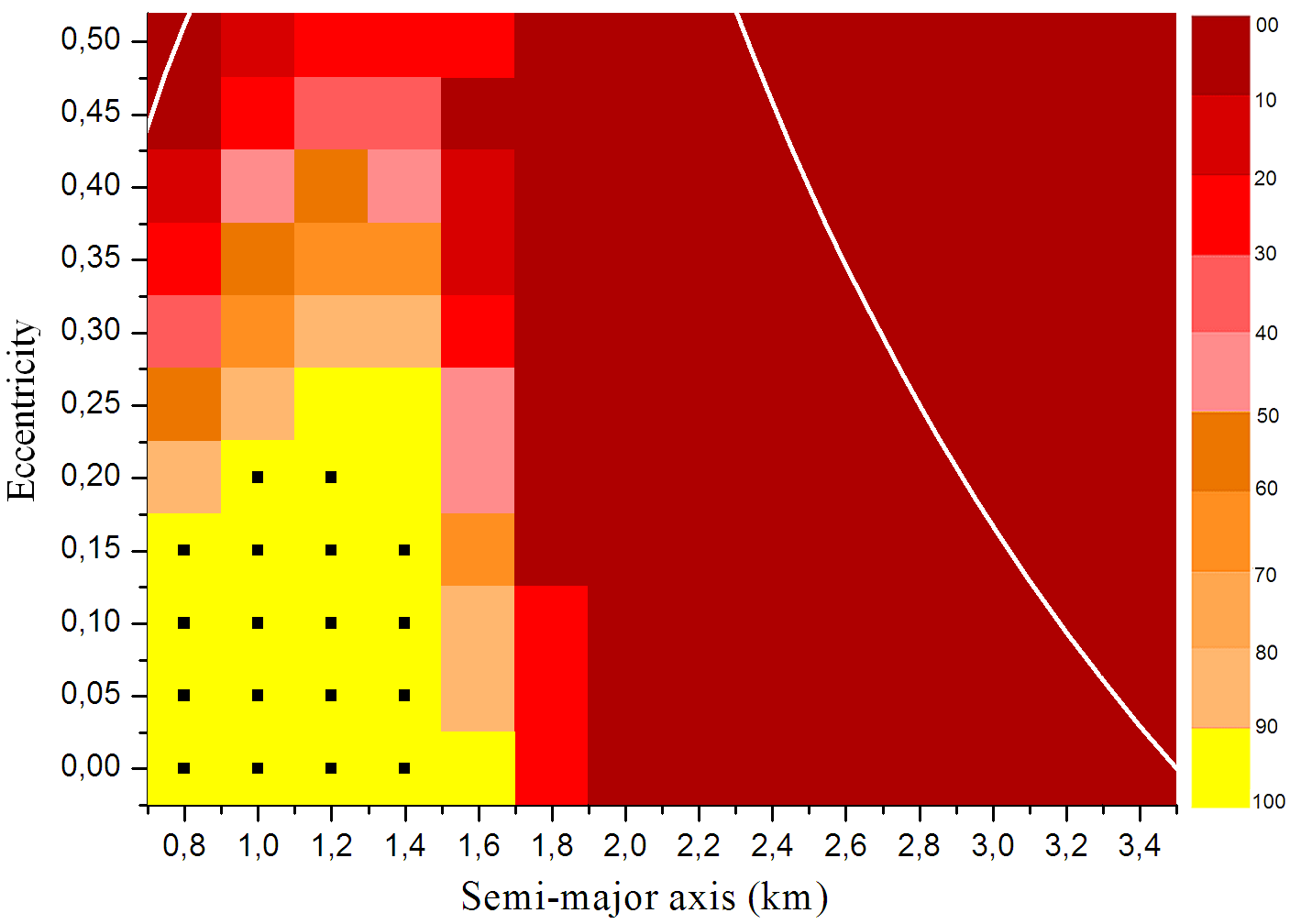}}\qquad
\subfigure[]{\includegraphics[height=5.2cm]{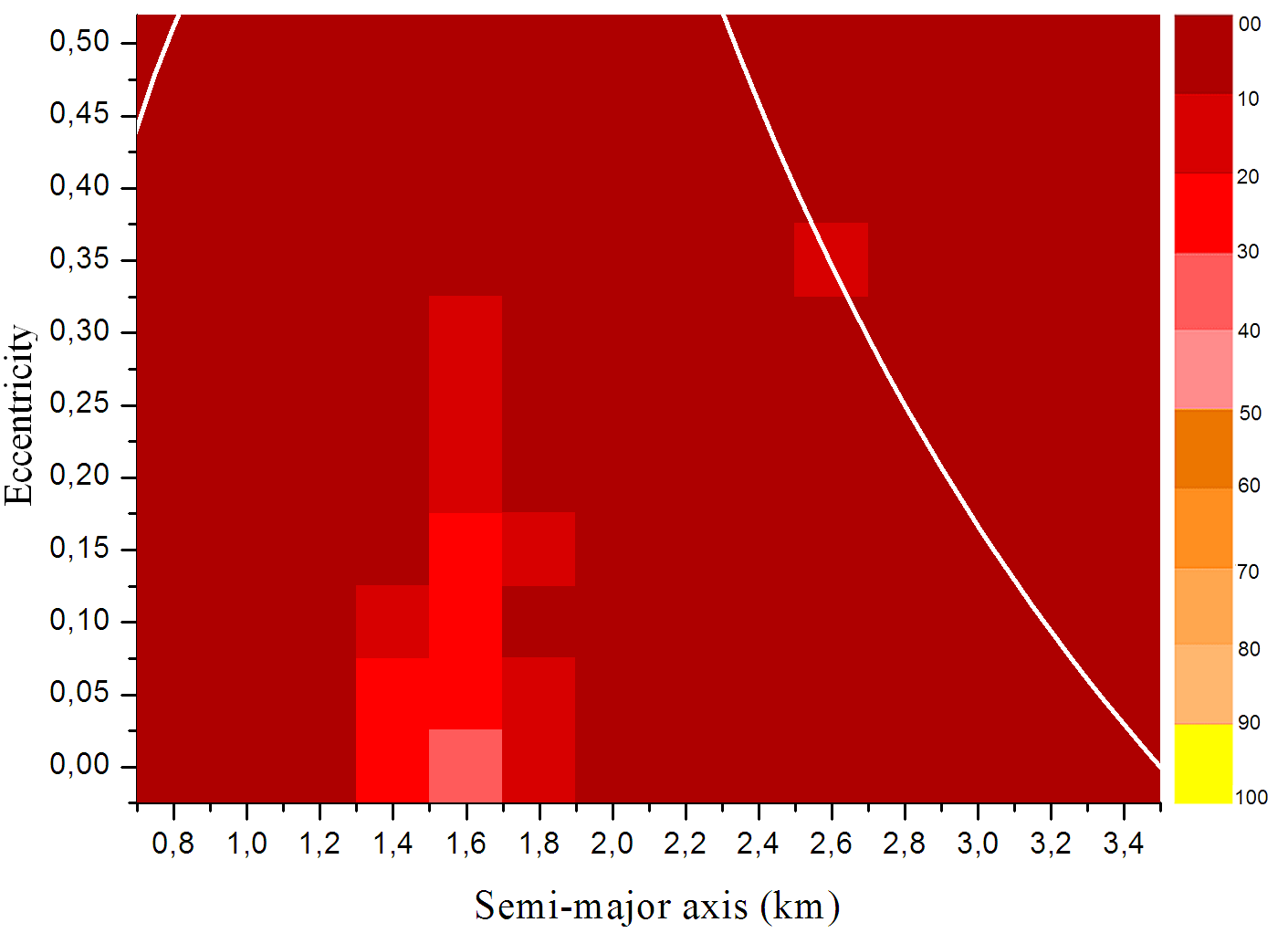}}}
\mbox{%
\subfigure[]{\includegraphics[height=5.2cm]{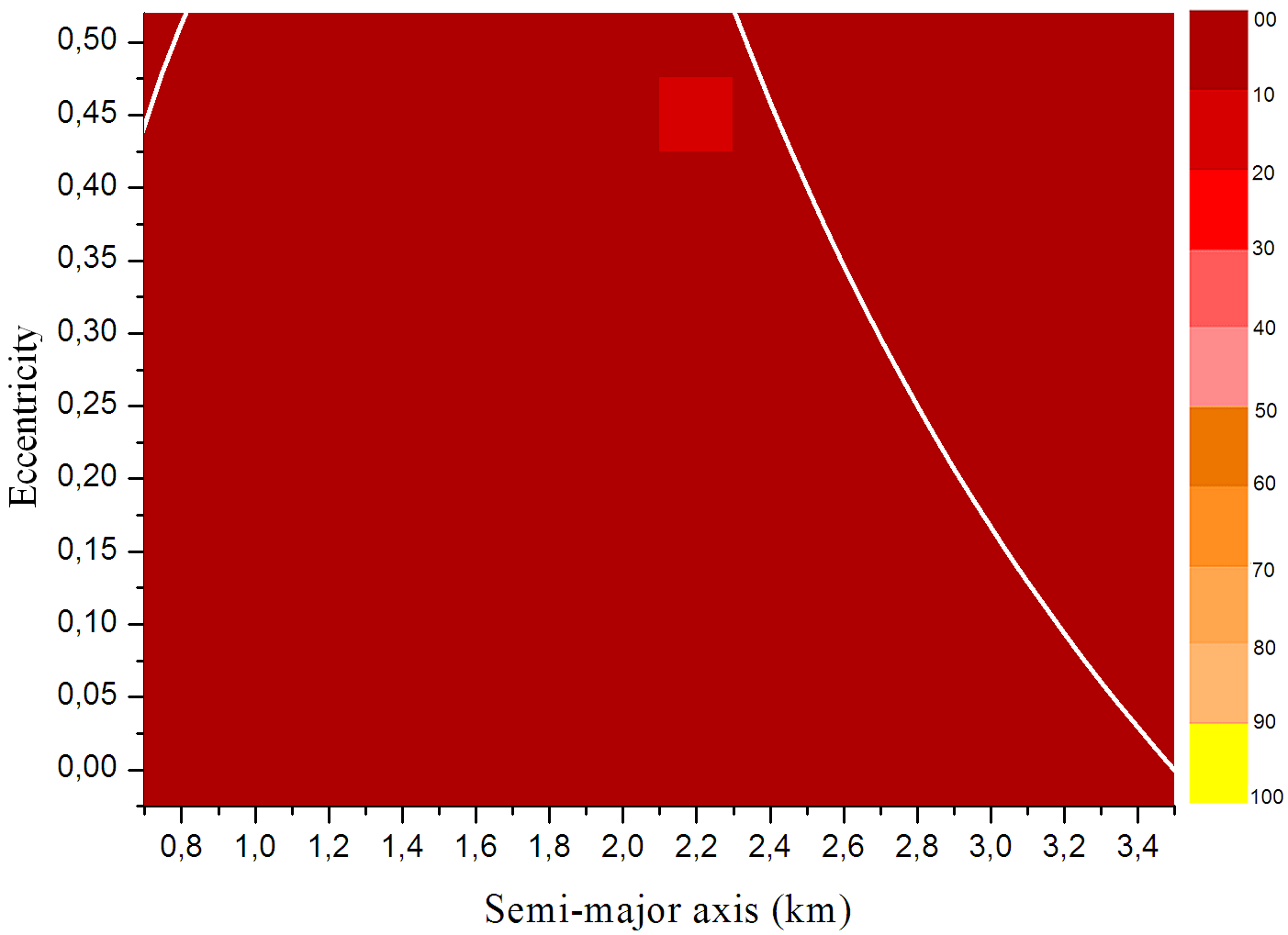}}\qquad
\subfigure[]{\includegraphics[height=5.2cm]{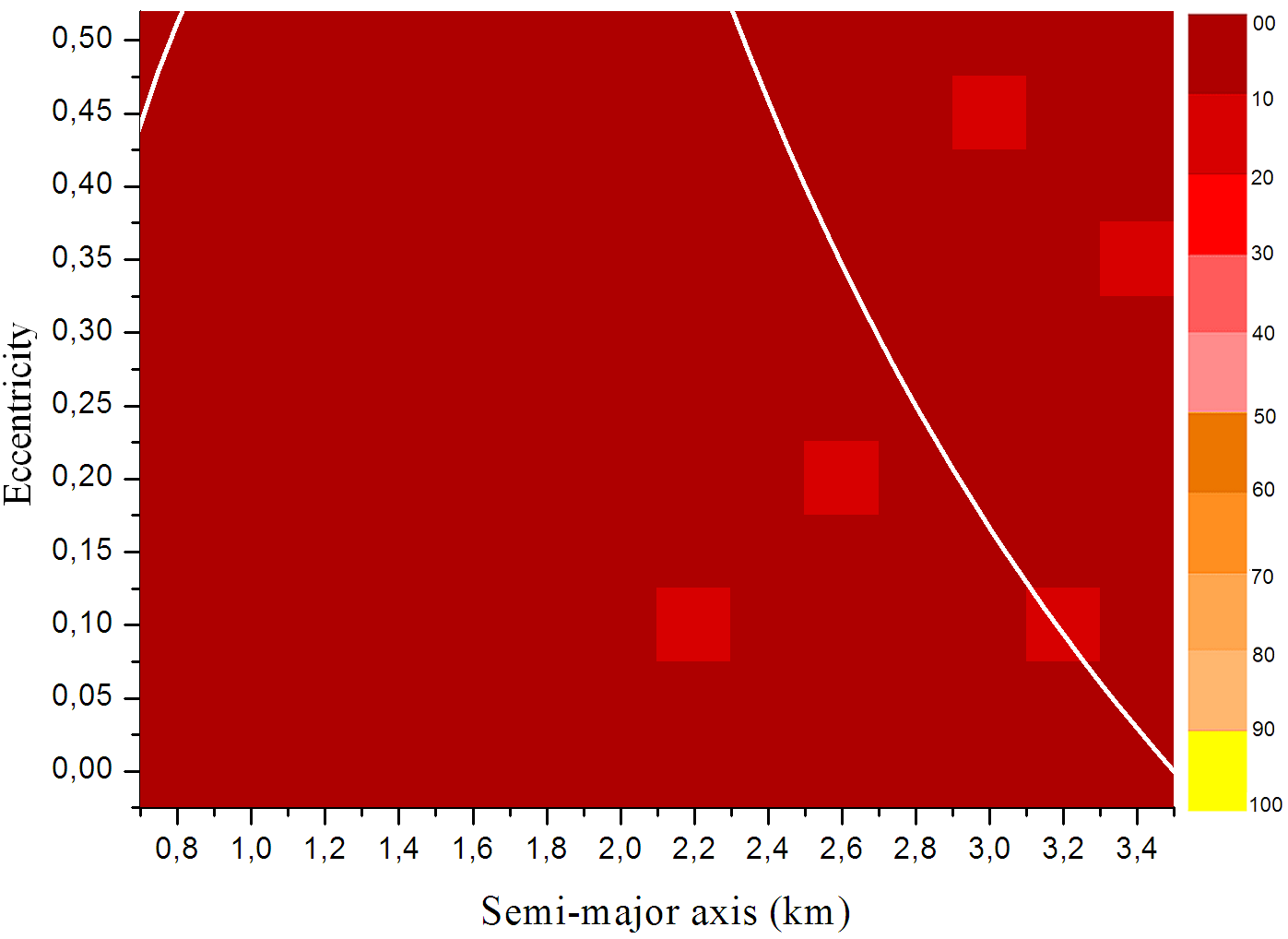}}}
\caption{Diagram of stability of region $3$, for a time span of 2 years. a) $I=15.0^{\circ}$, b) $I=30.0^{\circ}$, c) $I=45.0^{\circ}$, 
d) $I=60.0^{\circ}$, e) $I=75.0^{\circ}$, f) $I=90.0^{\circ}$. The white and blue lines indicate the limits of the region.The yellow boxes marked with the small
 black point indicate the cases of 100\% of survival.}
\label{fig_reg3}
\end{figure*}

\subsection{Region 1}

As in the planar case, the particles are orbiting Alpha with $1.4\leq a \leq 3.2$ (km) taken every $0.2$ km, $0.0 \leq e \leq 0.50$ taken
 every $0.05$, random values  for $f, \omega, \Omega$ and now with a variation on inclination:  $15.0^{\circ} \leq I \leq 90.0^{\circ}$ taken every $15.0^{\circ}$.

The regions of stability found, for the inclined prograde case in region 1, are presented on Figure \ref{fig_incl_re1}. Again a 
diagram of semi-major axis versus eccentricity was adopted. The limits of the region marked by the white lines are the same as discussed in subsection \ref{subsec_stabregion1}.

The region of stability found in region 1, for the planar case, remains for the
cases with inclination, until the critical value $I=60.0^{\circ}$, when the stable region decreases significantly. 
This is a phenomenon attributed to the Kozai mechanism \citep{b21}. In region 1, the particles with $I=60.0^{\circ}$ have an inclination relative to the 
orbital plane of Gamma of approximately $46^{\circ}$. Therefore, such particles have
exceed the \textit{critical angle of Kozai} given by $I_{crit}\approx39.2^{\circ}$. A known effect of the Kozai mechanism is generate oscillations of the eccentricity 
and of the mutual inclinations. These oscillations increase the probability of close encounters and collisions, and leads to the observed instability for larger 
values of inclination in region 1. An exception occurs for $I=90.0^{\circ}$ where it is observable a slight increase of the stable region. 

\cite{b26} presented a discussion on the role of the Kozai mechanism in binary NEAs. They considered a system composed by the Sun and by 
a binary system. In their model the secondary body orbits on the equatorial plane of the primary. The pole orientation of the primary is such that the 
relative inclination between the mutual
orbit of components of the binary and the heliocentric orbit of the binary system is in the interval $39.2^{\circ}<I<140.8^{\circ}$. The Sun is the perturber and they analyzed the 
perturbations suffered by the secondary body due to the Kozai mechanism and how the $J_{2}$ of the primary affects this problem. They found that 
for closely-separated binary the Kozai cycles can not occur. They showed that the precession of the pericentre of the secondary body due to the oblateness
of the primary is enough to suppress the effects due to the Kozai mechanism, even for very small $J_{2}$ values.

Considering their results, it would be expected for our case that the particles in region 1, placed close to Alpha (less than 3 radii of the primary), 
would suffer the same kind of phenomenon - i.e, the pericentre precession of the particles due to the oblateness of Alpha would suppress the effects due to the Kozai mechanism. 
Nevertheless, we have found that the particles in such region present the expected Kozai behavior, such as libration of the argument of 
pericentre and the increase of the eccentricity while the inclination decreases. Only the particles really close to Alpha do not 
present such Kozai behavior i.e, their argument of pericentre is circulating and the eccentricity and inclination suffer a small variation, leading to stability (see figure \ref{fig_incl_re1}d
for $a=1.4$ km and $a=1.6$ km, for instance). 

Thus, as discussed by \cite{b26}, we also found that the primary oblateness is capable to suppress the action of the Kozai mechanism, and 
that this depends on the distance from the perturbed body to the primary. However, the distances found for which each regime dominates is different from the distances found in the cited work.
This is mainly due to the fact that in our system the particles are not placed in the equatorial orbital plane of the primary, as in their model.

Analyzing the equation that gives the precession rate of the argument of pericentre due to $J_{2}$ \citep{b27}, we see that the maximum value occurs when the inclination of the perturbed body relative 
to the equatorial plane is equal to zero. When the perturbed body has a relative inclination different from zero, the precession rate of the argument of the pericentre decreases, 
and so decreases the efficiency  of the oblateness of the primary body to suppress the Kozai mechanism. 
As a consequence, the particles have to be closer to the primary in order to present stability, and a little farther particle will be unstable.  
That is what we observe for the particles in the internal region of the triple system 2001 SN263.  
They are not placed in the equatorial plane, and so they are not affected by $J_{2}$ as much as they would be affected if they were in a non-inclined orbit.

Analysis of the diagrams on Figure \ref{fig_incl_re1} also reveal the rise of a gap on the stable region as the inclination increases, located at $a=1.8$ km. Such behavior
has the characteristics of a resonance and is discussed in section \ref{sec_res}.

\subsection{Region 2}

As in the planar case, the particles are orbiting Alpha with semi-major axis $4.5 \leq a \leq 13.5$ (km) taken every $0.2$ km, $0.0 \leq e \leq 0.50$ 
taken every $0.05$, but now with a variation on inclination: $15.0^{\circ} \leq I \leq 90.0^{\circ}$ taken every $15.0^{\circ}$. The other orbital elements were chosen in the
same way as before. 

The regions of stability found for the inclined prograde case in region 2 are presented on Figure \ref{fig_incl_re2}. 
Comparisons between those diagrams and the diagram for the planar case (Figure \ref{fig_region2}) show a similar behavior, with distinct stable regions at the middle 
of the region and with a gap approximately at the center indicating the presence of resonance, discussed in section \ref{sec_res}.

From the same diagrams we see that, similar to observed for region 1, there is no stable region for $I \ge 60,0^{\circ}$, an indicative of Kozai mechanism 
that increases the eccentricities of the particles until they reach the collision or the ejection limit.

\subsection{Region 3}

As in the planar case, in region 3 the particles are orbiting Beta with semi-major axis $0.8 \leq a \leq 3.4$ (km) taken every $0.2$ km, with a variation on 
inclination: $15.0^{\circ} \leq I \leq 90.0^{\circ}$ taken every $15.0^{\circ}$. The other orbital elements were chosen in the
same way as before. 

The diagrams of semi-major axis versus eccentricity on Figure \ref{fig_reg3} show that, for particles in region 3 with inclination going from $I=0.0^{\circ}$
until $I=45.0^{\circ}$, the stable region is approximately the same, with particles really close to Beta and even with high eccentricities.

Similar to what happens with the particles on Region 2, there is no stable regions for inclinations higher than $60.0^{\circ}$, an effect caused by the Kozai mechanism.

\section{Internal regions: resonant motion}
\label{sec_res}

Analysis of the stability diagrams in Figures \ref{fig_incl_re1} and \ref{fig_incl_re2} reveal the rise of gaps in the internal regions 1 and 2 ($a \approx 1.8$ km for region 1
and $a \approx 8.0$ km for region 2). 
Such behavior has the characteristics of possible resonant motions between the particles on thin regions with Gamma or Beta. We have investigated this, and the discussions
are presented as follow.

\subsection{Resonance in region 1}
In order to characterize the resonant motion of the particles in region 1 with Gamma or Beta, we first performed a search for commensurabilities of mean motion. 
The mean motion of the particles and of the satellites were determined through the Equation \ref{eq_mean}, considering the oblateness of
the central body and the semi-major axis of the involved bodies \citep{b22}. 

\begin{equation}
n=\frac{G m{p}}{a^3}\left[1+\frac{3}{2}J_{2}\left(\frac{R_{p}}{a}\right)^2-\frac{15}{8}J_{4}\left(\frac{R_{p}}{a}\right)^4\right]
\label{eq_mean}
\end{equation}

Our analysis shows that the particles located at $a=1.8$ km are in a 3:1 commensurability of mean motion with Gamma. 
Knowing this relationship, it is necessary to determine the coefficients $j_{i}$ $\left(\sum_{j=1}^{6}j_{i}=0 \right)$, for which the resonant angle:
\begin{equation}
\varphi=j_{1}\lambda'+j_{2}\lambda+j_{3}\varpi'+j_{4}\varpi+j_{5}\Omega'+j_{6}\Omega
\label{eq_phi2}
\end{equation}
is librating, showing that the bodies are in resonance. In Equation \ref{eq_phi2}, $\lambda$ is the mean longitude, $\varpi$ is the longitude of pericentre and 
$\Omega$ is the longitude of the ascending node of the satellite, while $\lambda'$, $\varpi'$ and $\Omega'$ are related to the particle.

\begin{figure*}
\centering
\mbox{%
\subfigure[]{\includegraphics[height=5cm]{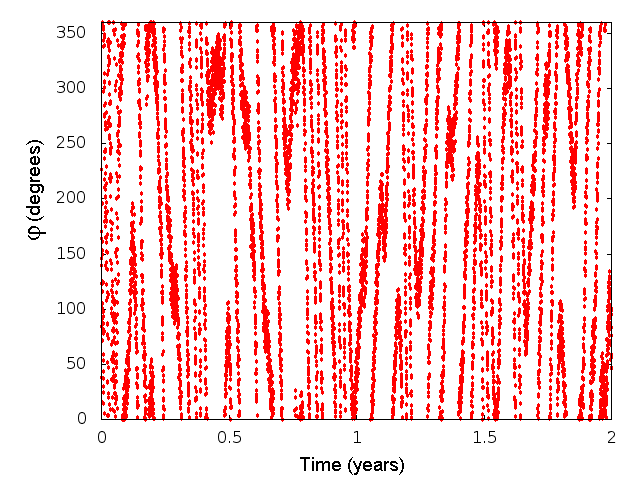}}\qquad
\subfigure[]{\includegraphics[height=5cm]{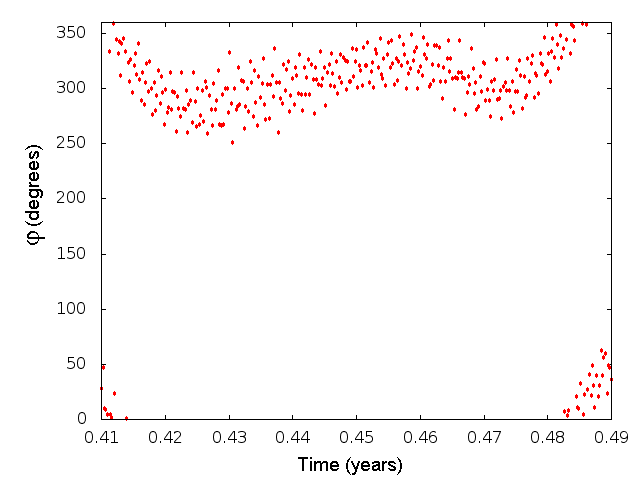}}}
\mbox{%
\subfigure[]{\includegraphics[height=5cm]{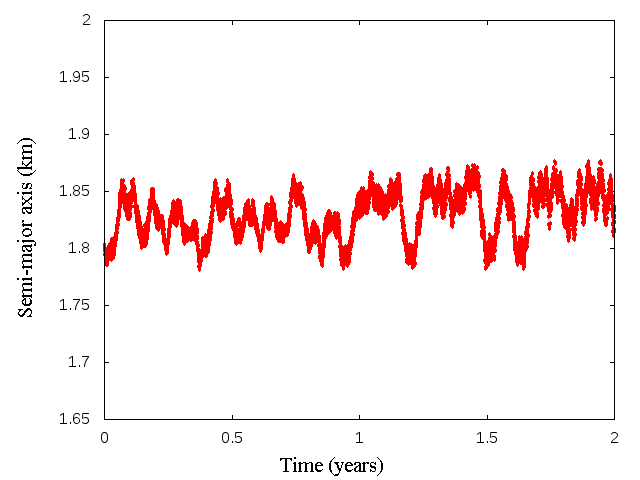}}\qquad
\subfigure[]{\includegraphics[height=5cm]{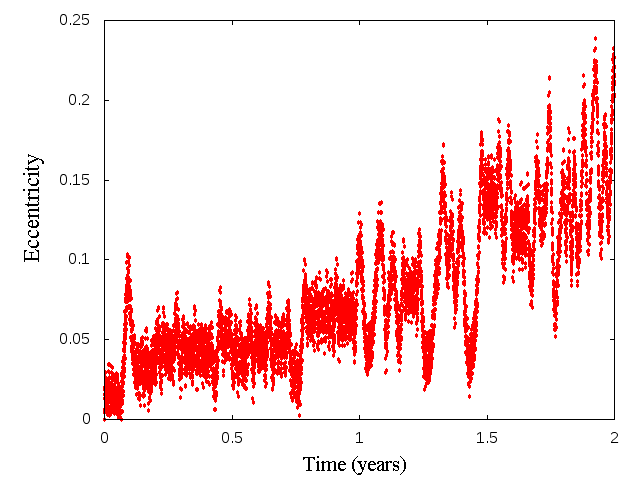}}}
\caption{Particle in region 1 with  $a=1,8$ km,  $e=0,0$ e $I=15,0^{\circ}$. a) Evolution of the resonant angle  $\varphi=3\lambda-\lambda'-\varpi'-\Omega'$ for $t=2$ years. 
b) Zoom of Figure (a) showing one of the libration regions. Time going from 0.41 until 0.49 years ($\approx42$ orbital periods of Gamma). c) Evolution of semi-major axis. d) Evolution of eccentricity.}
\label{fig_resonance1}
\end{figure*}

We found, considering a particle in region 1 with $a=1.8$ km, $e=0.0$ and $I=15.0^{\circ}$, that the resonant angle $\varphi=3\lambda-\lambda'-\varpi'-\Omega'$
presents an intermittent behavior (circulating and librating), as shown in Figure \ref{fig_resonance1} a and b. Therefore, particles at such neighborhood 
suffer the effects of a 3:1 resonance with Gamma. As a consequence, the particles on this region are perturbed in such way that the semi-major axis
remains almost constant while the eccentricity increases (Figures \ref{fig_resonance1} c and d), and then, the particles cross the collision-line of the region, 
giving rise to the observable instability.

\subsection{Resonances in region 2}

An investigation similar to the analysis done for region 1 show that particles in region 2,  with $a\approx8.0$ km, are in a 3:1 commensurability of mean motion with Beta.
Particles with $a\approx 7.9$ km, in the same region, are in a 1:3 commensurability of mean motion with Gamma.

\begin{figure*}
\mbox{%
\subfigure[]{\includegraphics[height=5cm]{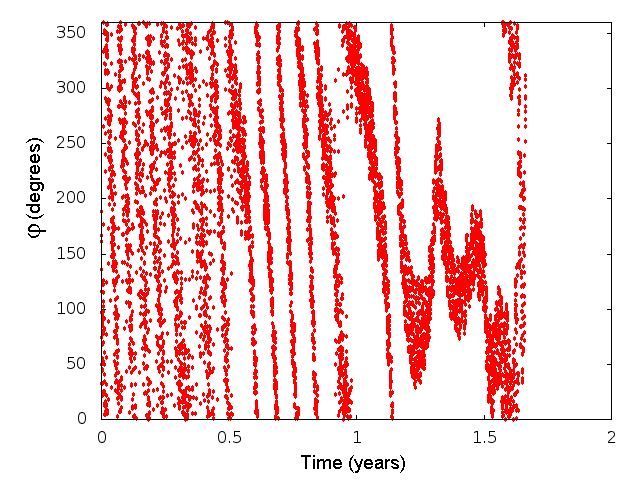}}
\subfigure[]{\includegraphics[height=5.0cm]{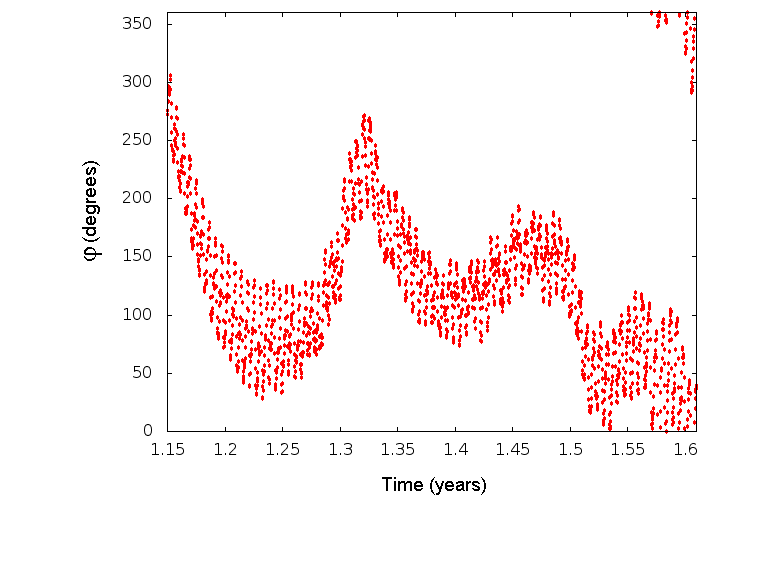}}}
\mbox{%
\subfigure[]{\includegraphics[height=5cm]{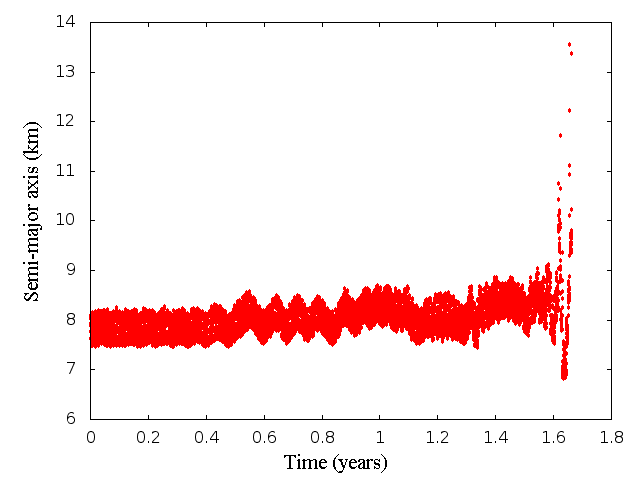}}
\subfigure[]{\includegraphics[height=5cm]{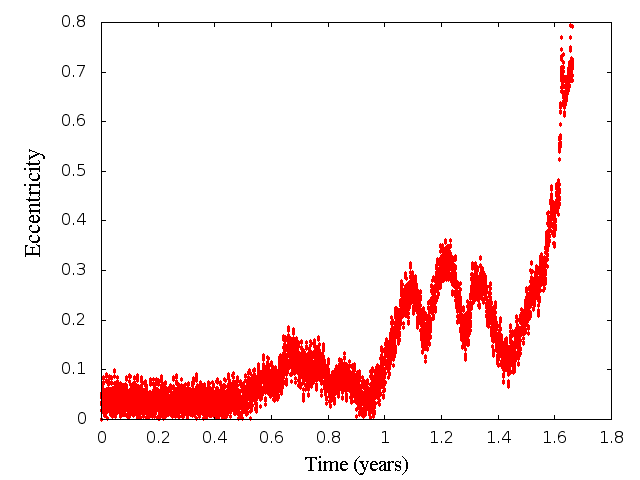}}}
\caption{Particle in region 2 with $a=7.9$ km, $e=0.0$ e $I=0.0^{\circ}$ that collides with Gamma at $t\approx1.6$ years.
a) Evolution of the resonant angle $\varphi=3\lambda-\lambda'-\varpi'-\Omega'$ for $t\approx 1.6$ years. 
b) Zoom of Figure (a) showing the large libration region. Time going from 1.35 until 1.5 years ($\approx9$ orbital periods of Beta). c) Evolution of semi-major axis. 
d) Evolution of eccentricity.}
\label{fig_res_beta}
\end{figure*}

\begin{figure*}
\mbox{%
\subfigure[]{\includegraphics[height=5.0cm]{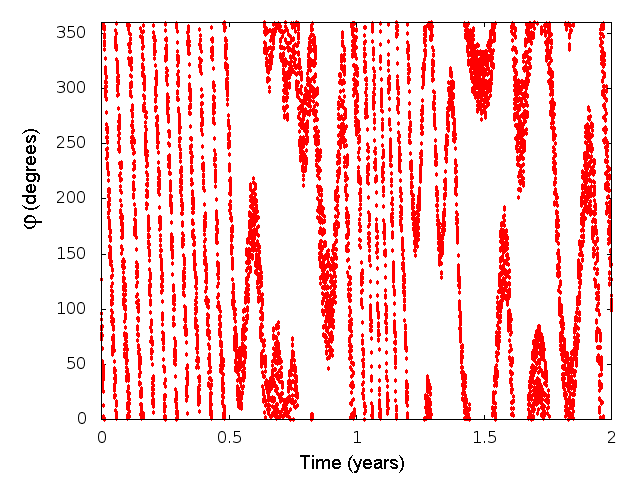}}
\subfigure[]{\includegraphics[height=5.0cm]{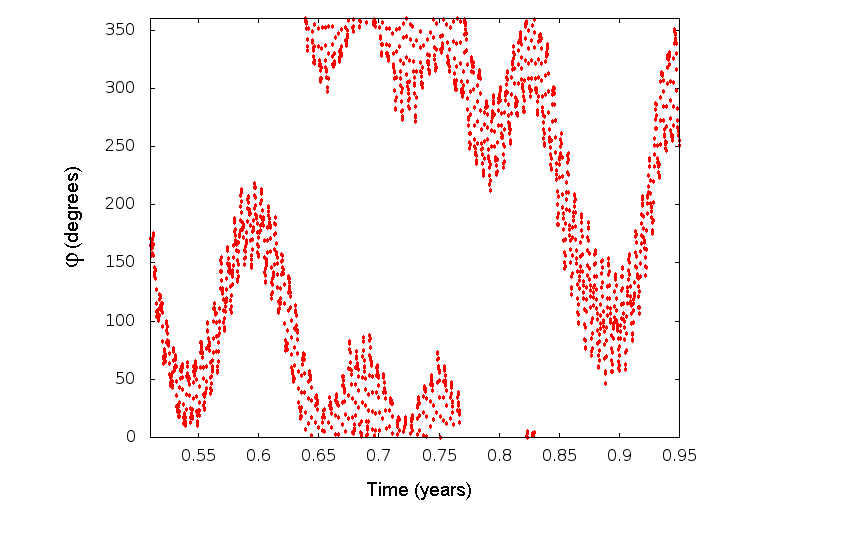}}}
\mbox{%
\hspace{-1.0cm}
\subfigure[]{\includegraphics[height=5.0cm]{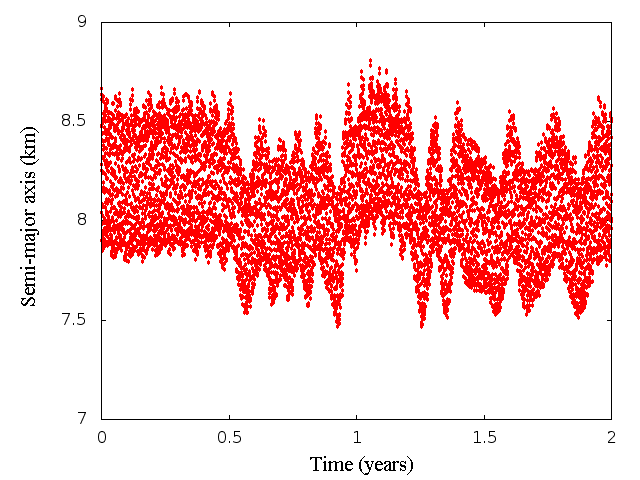}}
\subfigure[]{\includegraphics[height=5.0cm]{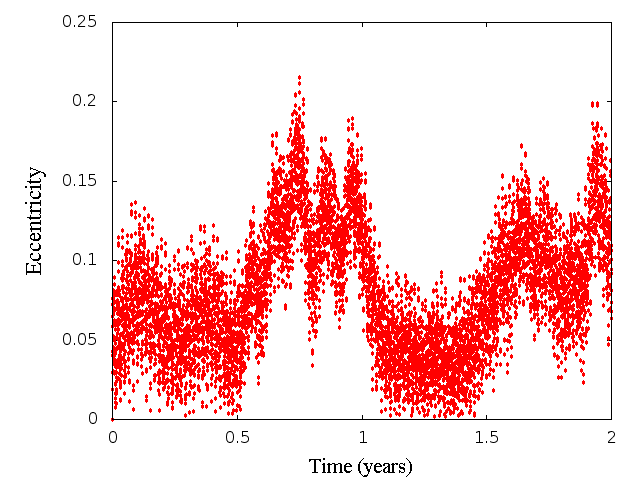}}}
\caption{Particle in region 2 with $a=7.9$ km, $e=0.0$ e $I=0.00^{\circ}$.
a) Evolution of the resonant angle $\varphi=3\lambda'-\lambda-\varpi-\Omega$ for $t=2$ years. 
b) Zoom of Figure (a) showing one of the libration regions. Time going from 0.85 until 0.92 years ($\approx37$ orbital periods of Gamma). c) Evolution of semi-major axis.
d) Evolution of eccentricity.}
\label{fig_res_gama}
\end{figure*}

For the resonance with Beta, we found that the resonant angle $\varphi=3\lambda-\lambda'-\varpi'-\Omega'$ presents an intermittent behavior 
(circulating and librating), as shown in Figures \ref{fig_res_beta} a and b. Therefore, particles in such neighborhood 
suffer the effects of a 3:1 resonance with Beta. As a consequence, the particles on this region are perturbed in such way that the semi-major axis
remains almost constant while the eccentricity increases (Figures \ref{fig_res_beta} c and d), and then the particles cross the collision-line of the region, 
giving rise to the observable instability. Similar behavior was found for the resonance 1:3 with Gamma, for the resonant angle $\varphi=3\lambda'-\lambda-\varpi-\Omega$, as 
can be seen in Figure \ref{fig_res_gama}.

\section{The external region}

In the previous sections we characterized the stability of the internal regions of the triple system 2001 SN263, for the planar and inclined prograde cases. Now we extend our
analysis to the external region. 

The methodology adopted is the same as before, i.e., we performed numerical integrations for thousand of particles orbiting Alpha, but now,
with orbits external to the system. To obey such condition the particles must be more distant to Alpha than Beta (lower limit). The upper limit is defined taking into account
the approximated
Hill's radius of the whole system. Considering a body whose mass is equal to the sum of the mass of Alpha, Beta and Gamma, and the orbit of the system relating to the Sun,
we have calculated that the Hill's radius of such hypothetical body is $R_{Hill}\approx180.0$ km at perihelion of the orbit and $R_{Hill}\approx500.0$ km at aphelion.
According to \cite{b18} the limit of stability is about $0.5$ Hill's radius for prograde cases. Thus, the value for the upper limit of the external region  was chosen 
to be $d=90.0$ km ($\approx0.5 R_{Hill}$ at perihelion, when the system is more perturbed). Beyond the distance $d$, the particles are considered ejected. The collisions with any of the bodies are also considered.

Based on such limits we have the initial conditions for the particles in the external region, being: $20.0 \leq a \leq 90.0$ (km), taken every $1.0$ km, $0.0 \leq e \leq 0.50$ 
taken every $0.05$, and 100 particles for each pair $(a \times e)$ with random values  for $f, \omega, \Omega$. The combination of these initial conditions resulted in   
$78.100$ particles placed at such region.

We have performed numerical integration  using the Gauss-Radau integrator for a time span of $2$ years of the problem composed by those particles and by seven massive bodies:
Sun, the planets Earth, Mars and Jupiter, and the triple system of asteroid, considering again the planar case ($I=0.0^{\circ}$) and the inclined prograde cases 
$15.0^{\circ} \leq I \leq 90.0^{\circ}$. The results are presented in the next subsections.

\subsection{Planar case}

The regions of stability found for the external region, for the planar case, is presented on the first diagram of Figure \ref{fig_r_ext}. As before, it was considered a grid 
of semi-major axis versus eccentricity, and each one of the small ``boxes'' hold the information of $100$ particles that share the same initial values for $a$ and $e$. The 
coded color indicates the percentage of survivors, as it was previously defined.

The limits of the region are represented on the diagram by the white lines. On the left is the collision-line with Beta, which can also be understood as the limit
between the internal and external regions, according to their definition. Considering that the limit of the internal region is given by $d=16.633$ km, and being $q=a(1-e)$ 
the distance of the particles to Alpha at the pericentre, then, the relation $a=d/(1-e)$ numerically give the limit for what the particles cross the orbit of Beta. 
On the right is the ejection-line, denoting the limit from which the apocenter $(Q)$ of the orbit of the particle is greater than the ejection distance $d=90.0$ km.
Being $Q=a(1+e)$, then, the relation $a = d/(1+e)$, with $0.0\leq e \leq 0.5$, gives the limit of ejection for the external region.

From such diagram of stability we see that the unstable regions are found beyond the limit of the ejection-line, and at the neighborhood of Beta, caused most by the collisions
of the particles with this body, as can be seen in Figure \ref{fig_estatistics}. According to the same figure, about $30\%$ of the particles collide or are ejected.

Approximately $27\%$ of the particles are in the region where $100\%$ of the particles survived $2$ years (yellow region marked with the small black point), hence,
the stable region is predominant in the external region.


\subsection{Inclined prograde case}

The initial conditions, the number of particles and the limits of the region are exactly the same for the
planar case, except by the inclination of the particles that was taken from  $15.0^{\circ}$ until $90.0^{\circ}$, in a fixed interval of $15^{\circ}$.

\begin{figure*}
\subfigure{\includegraphics[height=3.0cm]{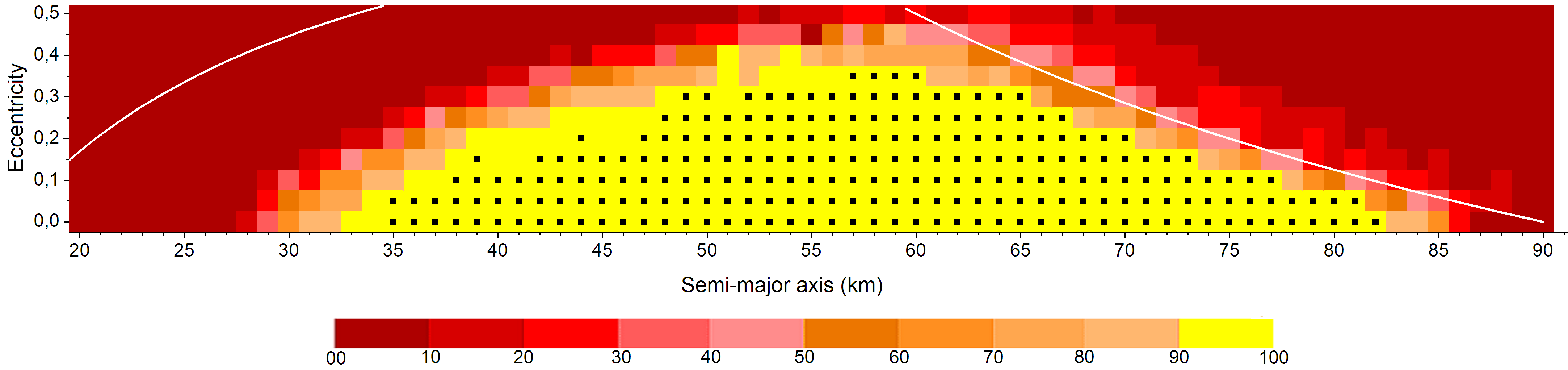}}
\subfigure{\includegraphics[height=3.0cm]{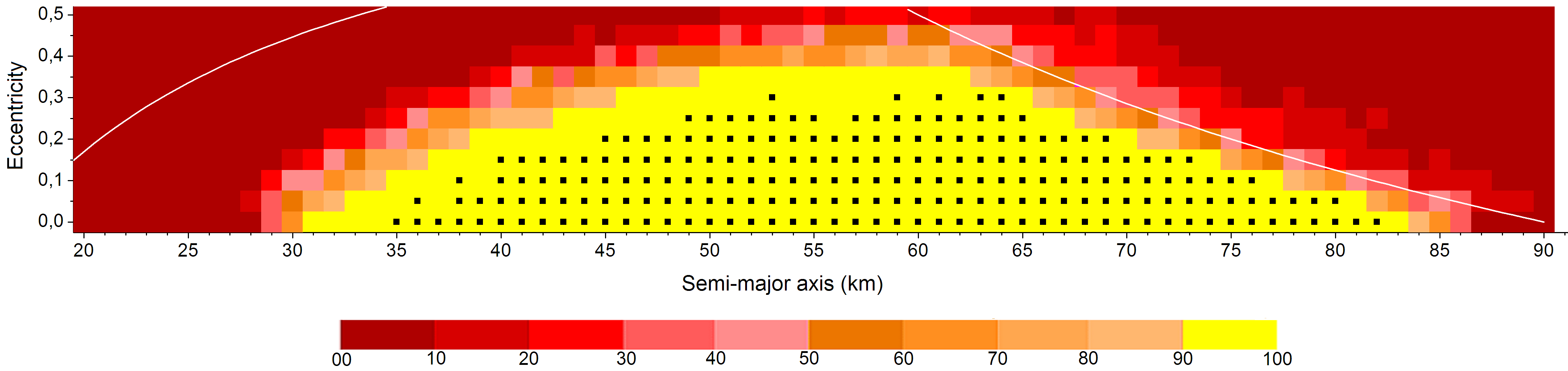}}
\subfigure{\includegraphics[height=3.0cm]{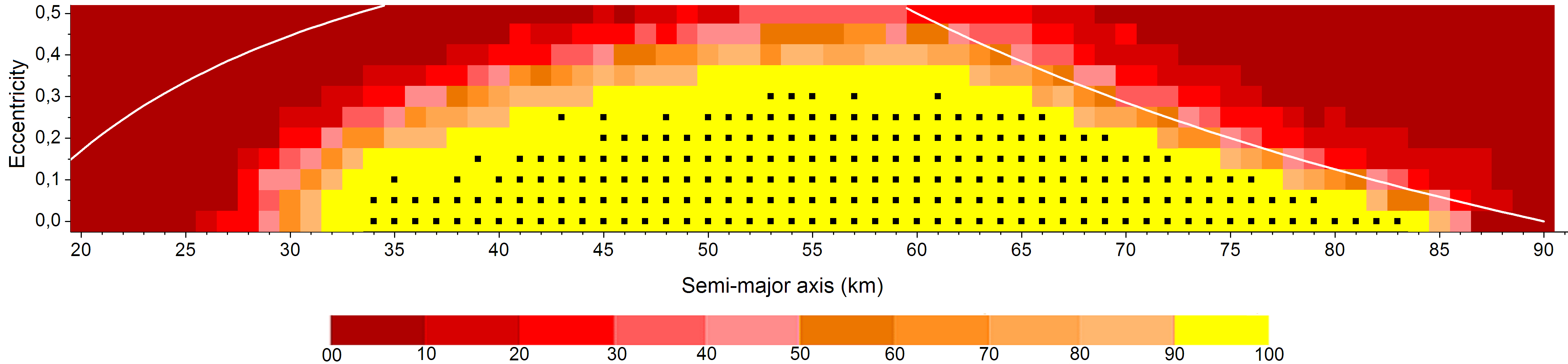}}
\subfigure{\includegraphics[height=3.0cm]{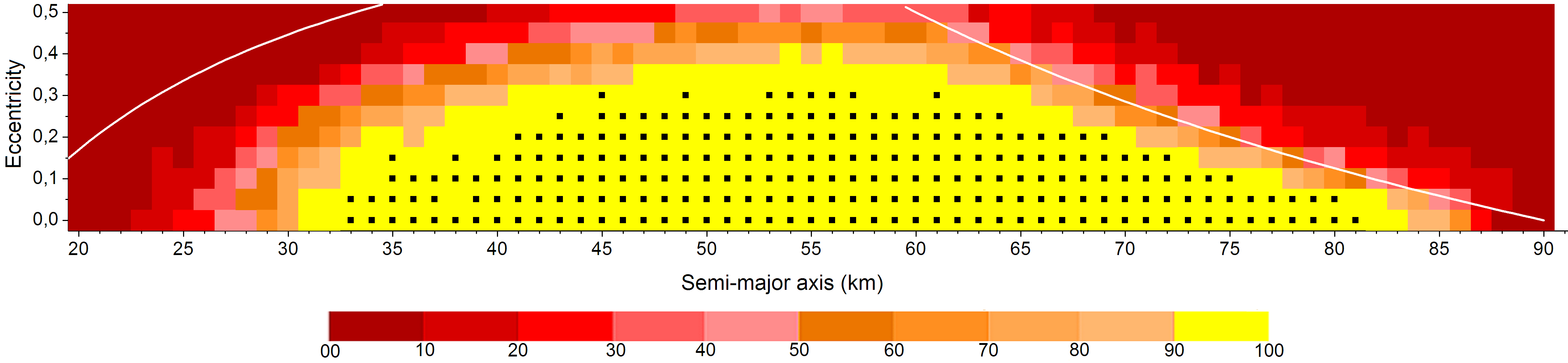}}
\subfigure{\includegraphics[height=3.0cm]{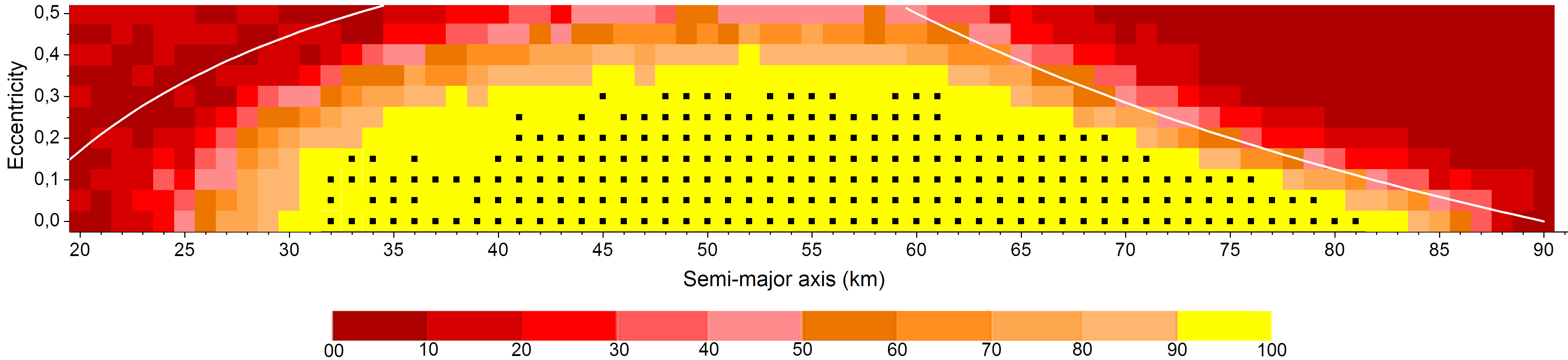}}
\subfigure{\includegraphics[height=3.0cm]{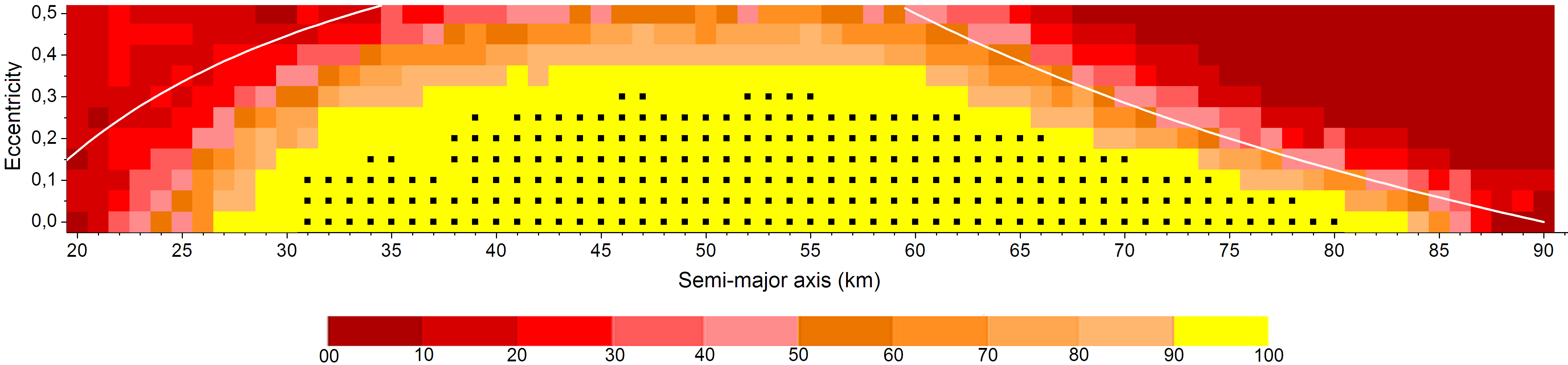}}
\subfigure{\includegraphics[height=3.1cm]{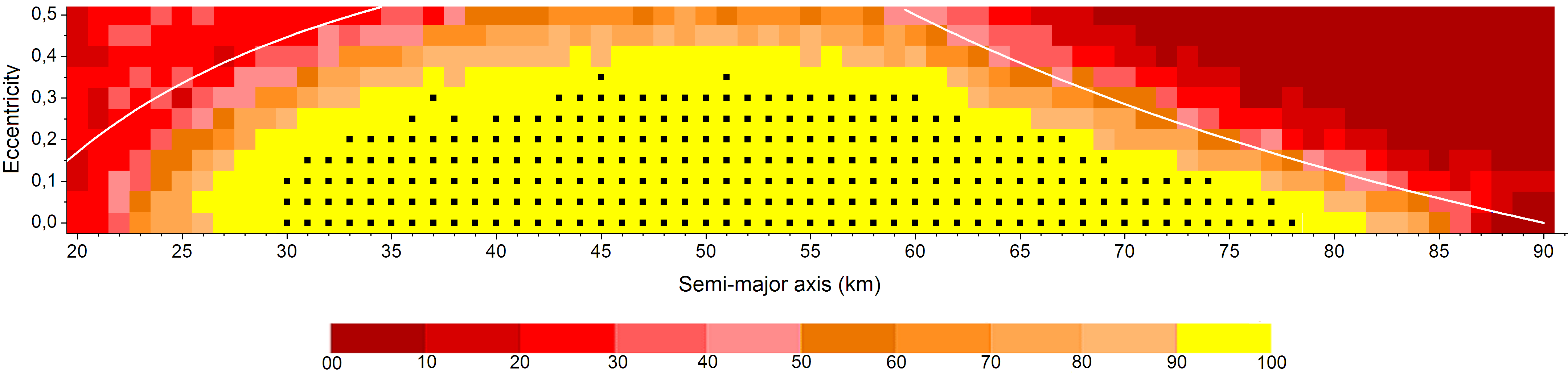}}
\caption{Diagram of stability of the external region for a time span of 2 years for $I=0,0^{\circ}$, $I=15,0^{\circ}, 30,0^{\circ},I=45,0^{\circ}, I=60,0^{\circ}, 
I=75,0^{\circ}$ and $I=90,0^{\circ}$, from top to bottom. The white line indicate the limits of the region.The yellow boxes marked with the small
 black point indicate the cases of $100\%$ of survival.}
\label{fig_r_ext}
\end{figure*}

The stable regions found for those conditions are presented in the diagrams of Figure \ref{fig_r_ext}. 

Comparing those diagrams with each other, and with the diagram for the 
planar case, we conclude that the stable regions are almost the same, and so, that the variation of inclination do not affect the stability of the external region of the triple
system.

\section{Long-term stability - planar and circular case}

Here we present a study on the long-term stability for the planar and circular case, in the internal region. 
We have restricted our analysis to this special case because it is more interesting for a spacecraft mission exploration point of view, and also
 due to the substantial computational processing time required 
to perform the long-term stability integrations for the eccentric and inclined cases as well as for the particles placed in the external region.

We have performed numerical integrations considering the particles with $e=0.0$, $I=0.0^{\circ}$ from regions 1,2 and 3, belonging to the set of initial 
conditions that resulted in $100\%$ of survival (yellow region marked with the small black points), for a time span of 2000 years. 

The period of 2000 years corresponds to approximately 700 orbital periods of Alpha, 106.000 orbital periods of Beta, and, a million of orbital periods of Gamma. 
Throughout this period there were no close encounters (within 1 Hill's radius) between the system and the planets Earth and Mars. Nevertheless, the period of 2000 years 
is sufficient to capture the effects of secular perturbations experienced by the system.

\begin{figure}
\subfigure[]{\includegraphics[height=3.3cm]{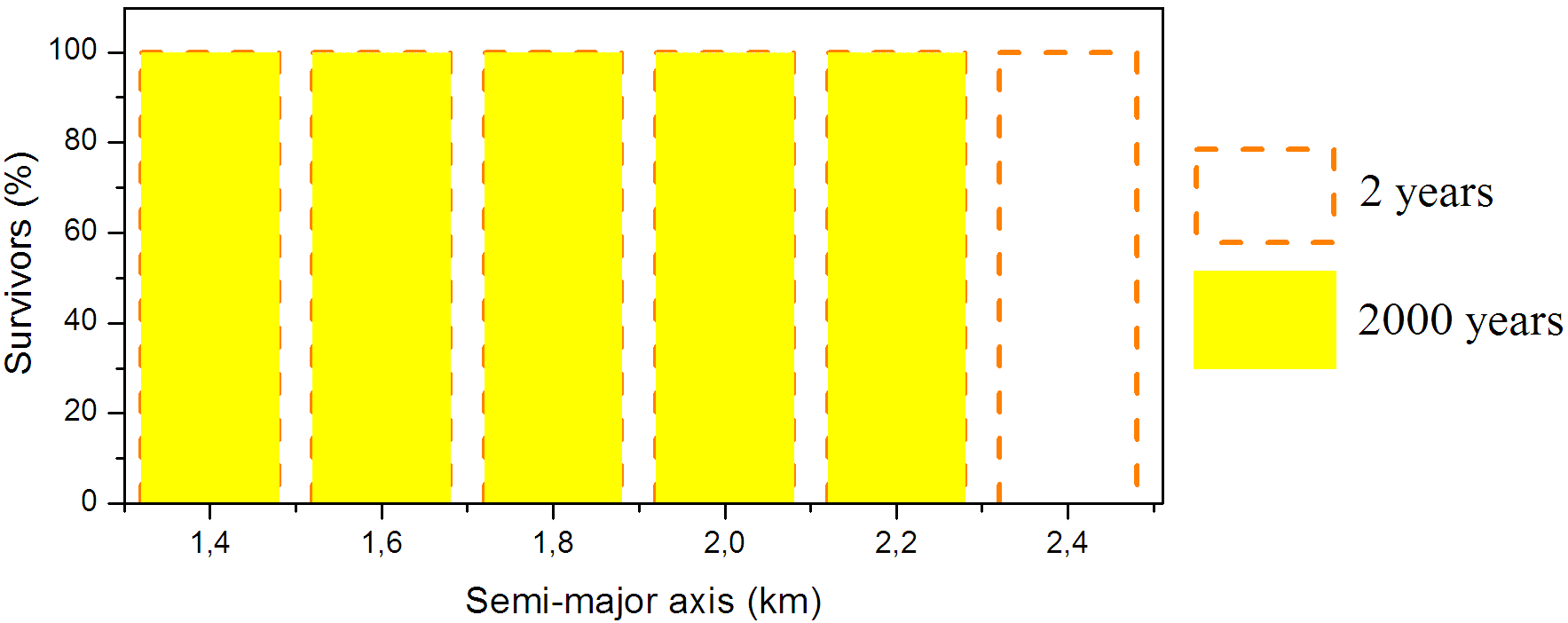}}
\subfigure[]{\includegraphics[height=3.3cm]{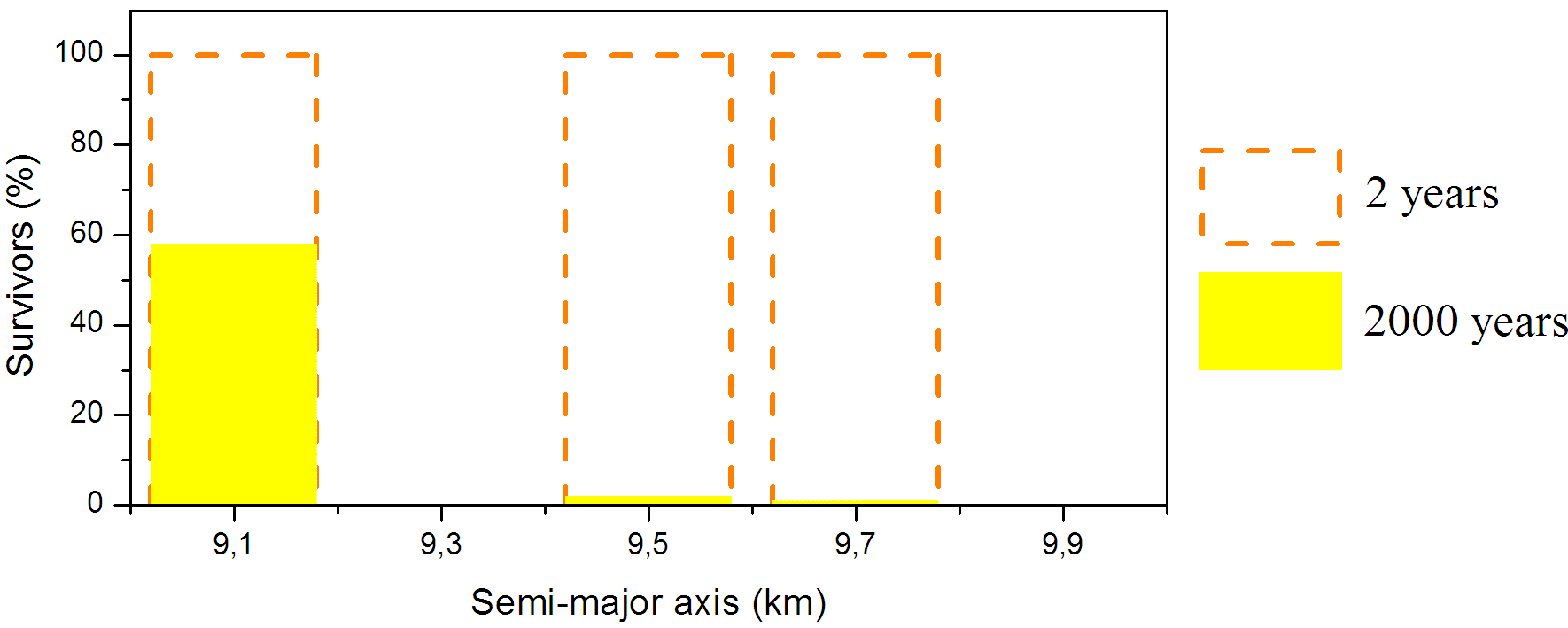}}
\subfigure[]{\includegraphics[height=3.3cm]{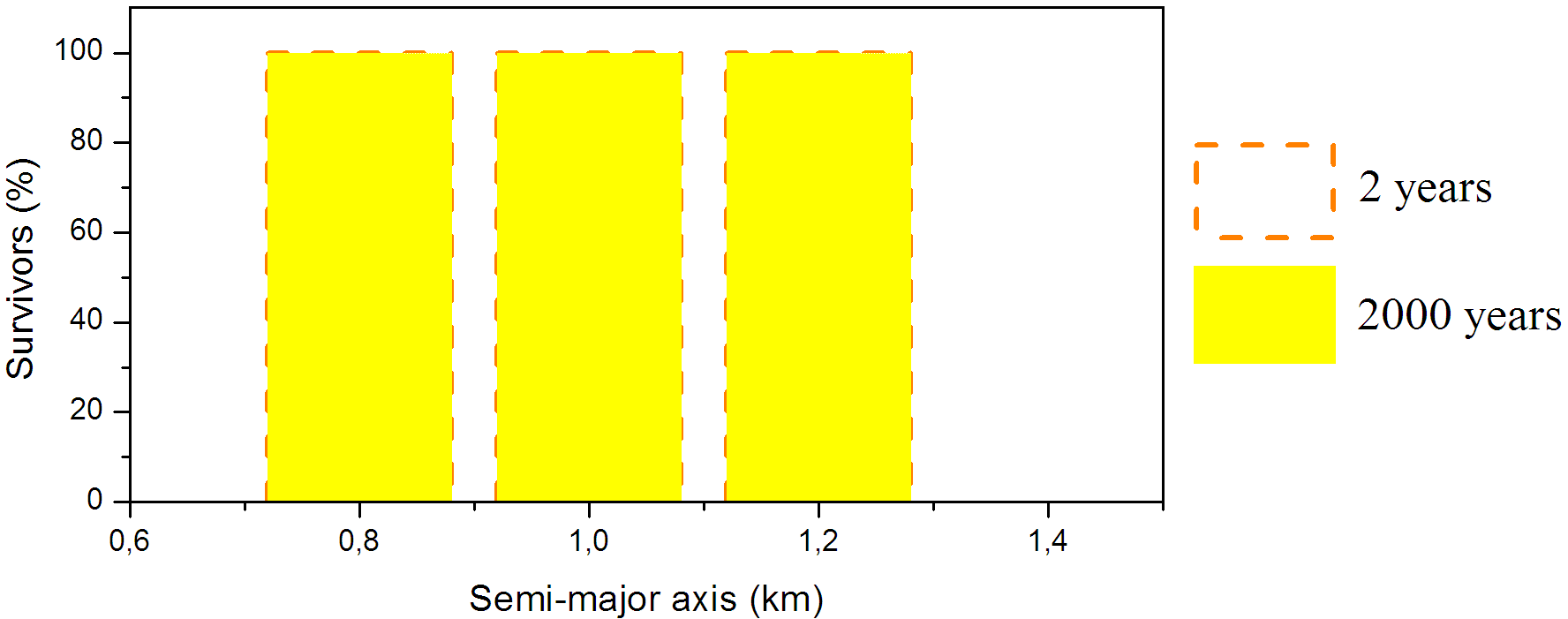}}
\caption{Percentage of survivors for each semi-major axis from a) region $1$, b) region $2$, and c) region $3$, corresponding to stable region, for the planar and circular case. 
The dashed orange line indicates the percentage of survivors for a time span of 2 years. The filled yellow column indicates the percentage of survivors for a 
time span of 2000 years.}
\label{fig_longterm}
\end{figure}

The results found for each one of the internal regions are presented on the diagrams of Figure \ref{fig_longterm}. In such diagrams we make a comparison between the percentage
of survivors after 2 years and after 2000 years, for each initial semi-major axis from region $1$, $2$ and $3$. The results are discussed separately.

\begin{itemize}
 \item Region 1
\end{itemize}

In region 1, the set of particles with $e=0.0$ and $I=0.0$, stable for 2 years, are those with semi-major axis going from 1.4 km until 2.4 km 
(see the diagram of stability in Figure \ref{fig_re1}).
From diagram in Figure \ref{fig_longterm}a, we see that the region of stability for 2000 years remains almost the same as that for 2 years.
Only the most external particles, with $a=2.4$ km, were removed ($100\%$ of collisions within 592 years).

\begin{itemize}
 \item Region 2
\end{itemize}

For region $2$, we have that only $3$ specific cases resulted in stability, i.e., $100\%$ of survival in 2 years, for the planar and circular case (see the diagram of
stability in Figure \ref{fig_region2}). 
The long-term integrations show that those stable cases no longer survive for 2000 years. From figure \ref{fig_longterm}b we see that the region of stability previously found
for $2$ years in region $2$, vanishes for the  long period.
$48\%$ of the particles with $a=9.1$ km are ejected from the system within $115$ years, while just $52\%$ of them survive 2000 years.
For particles with $a=9,3$ km and $a=9,5$ there were  $85\%$ of ejections and $15\%$ of collisions, and therefore, no particles survive 2000 years.

\begin{itemize}
 \item Region 3
\end{itemize}

In region 3, the set of particles with $e=0.0$ and $I=0.0$, stable for 2 years, are those with semi-major axis going from 0.8 km until 1.2 km 
(see diagram of stability in Figure \ref{fig_region3}).
For region $3$ we show that the region of stability for 2000 years remains the same as that found for 2 years (no collisions or ejections), 
as can be seen in Figure  \ref{fig_longterm}c. \\

\begin{figure}
\subfigure[]{\includegraphics[scale=0.041]{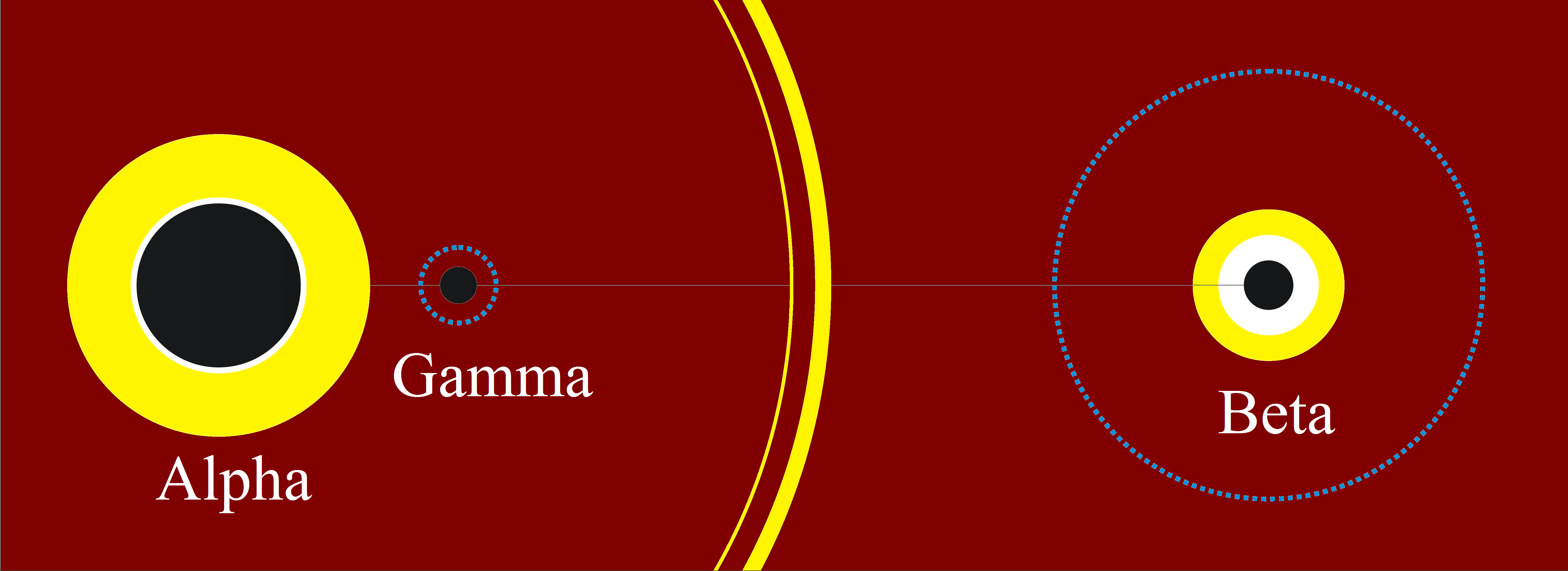}}
\subfigure[]{\includegraphics[scale=0.042]{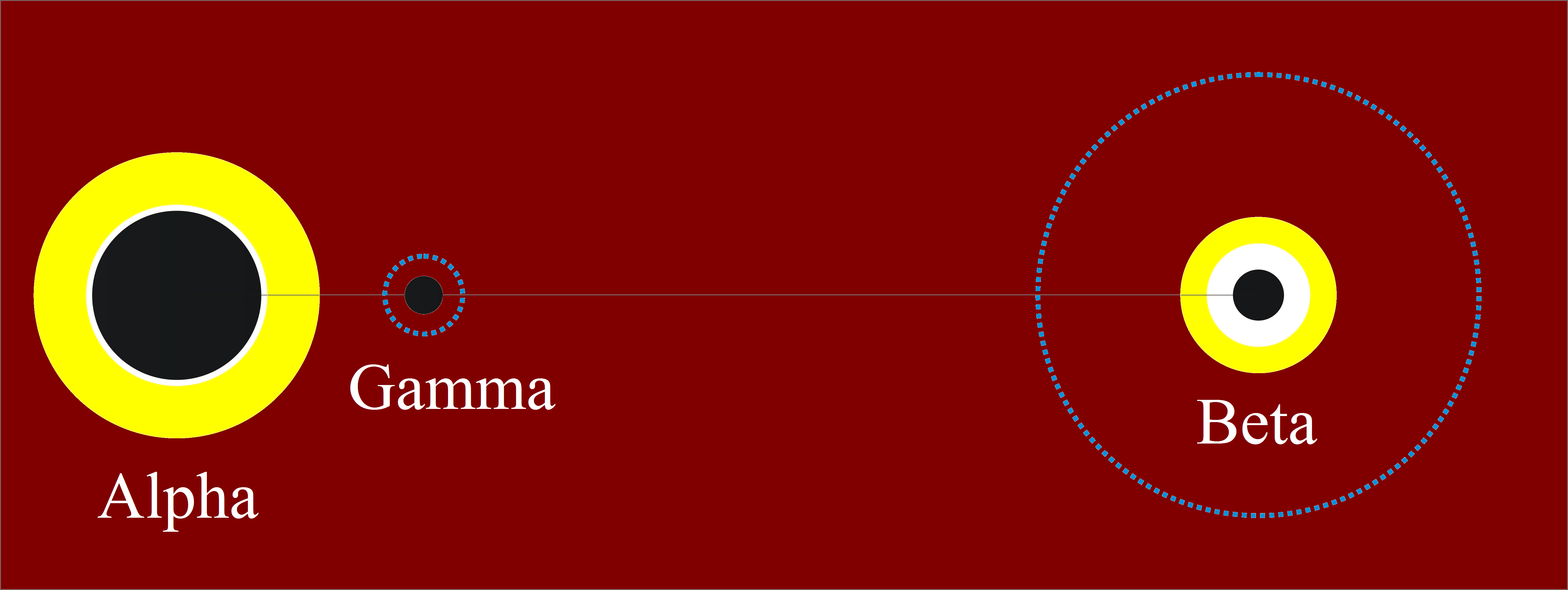}}
\caption{Representation of the stable (yellow) and unstable (red) regions in the internal region of the triple system 2001 SN263, considering the planar and circular case. 
a) Short-term stability (period of 2 years). b) Long-term stability (period of 2000 years).}
\label{fig_final}
\end{figure}

The regions of stability found for short and long periods can be better visualized in the diagrams presented in Figure \ref{fig_final}, where it was considered 
only the special case treated in the present section, i.e., particles with circular and planar orbits in the internal regions of the triple system.

Both diagrams clearly show that the stable regions are really close to Alpha and Beta. For short-term period (Figure \ref{fig_final}a) we have the tiny stable conditions in
region 2 (region between Gamma and Beta) that vanishes for the long-term period (Figure \ref{fig_final}b). That was the most significant observable change between the 
short and long period cases. Besides that we had the smooth decrease of the stability region in region 1 (region between Alpha and Gamma).

The results show that the system has a quick dispersion of particles, i.e., the observable unstable region for 2 years is almost the same for 2000 years.

The long-term stable regions (region yellow in Figure \ref{fig_final}b) are an indication of regions where possibles satellites or debris would be found within the system. 
On the other hand, the short-term unstable regions (region red in Figure \ref{fig_final}a) are an indicative of regions where such bodies would not be found. 

\section{Final comments}
\label{sec_final}

In this paper, we characterized the internal and external regions of stability and instability for the triple system of asteroid 2001 SN263, for a time span of $2$ years.
The region around the system was divided into four distinct regions (three of them internal to the system and one external). We have performed numerical 
integrations of systems composed by seven bodies: Sun, Earth, Mars, Jupiter and the three components of the system, and by thousands of particles randomly distributed 
within the demarcated regions, for the planar and inclined prograde cases.

For the planar case, it was shown that in the internal region the stable regions are very close to Alpha (region 1) and Beta (region 3). 
In the region among Beta and Gamma, here called region 2, we showed that the particles suffer the effects of resonances with Beta and Gamma. 
We identified that the particles in region 2 with $a\approx8.0$ km are in a 3:1 resonance
with Beta, and that particles with $a\approx7.9$ km, in the same region, are in a 1:3 resonance with Gamma. The presence of such resonances and the presence of Beta and Gamma 
surrounding the region turn the region 2 a unstable region. The stability in region 2 was found only for some specific values of initial semi-major axis ($a=9.1, 9.5, 9.7$ km), 
and with initial eccentricity $e=0.0$.

We also analyzed the inclined prograde case for the internal regions. They were considered particles with inclination relative to the equator of the central body
in the interval $15.0^{\circ}\leq I \leq 90.0^{\circ}$, taking every $15.0^{\circ}$.
For regions 2 and 3 we found that, until an inclination of $I=30.0^{\circ}$ for region 2, and an inclination of $I=45.0^{\circ}$ for region 3, 
the stable regions are similar to the regions found for the planar case. 
For inclinations higher than $I=45.0^{\circ}$, the stable regions vanish. With such inclinations, the particles have exceed the critical angle of Kozai 
given by $I_{crit}\approx39.2^{\circ}$. A known effect of the Kozai mechanism is generate oscillations of the eccentricity 
and of the mutual inclinations. These oscillations increase the probability of close encounters and collisions, and lead to the observed instability.
The effects due to the Kozai mechanism is also observable in region 1. The stable region decreases significantly for particles with $I\geq60.0^{\circ}$ (an exception for $I=90.0^{\circ}$, when there is
a slight increase of the stable region). The results for region 1 also show that, with the increase of inclination, the effect of the resonant motion between the particles and Gamma
becomes observable. We identified that the particles in this region with $a\approx1.8$ km are in a 3:1 resonance with Gamma. 

For the external region we found that the stable region is predominant. It was shown that the instability is found only on the neighborhood of the limits of the region, i.e., 
on the vicinity of asteroid Beta (high collision probability) and beyond the border of the ejection-line. 
Inside those limits we found a significant stable region, where a large number of particles survives for $2$ years.
The analyze of the inclined prograde cases for the external region shows that this region is not affected by the variation of inclination of the particles. 

We also made an analysis of the long-term period stability for the internal regions, considering the planar and circular cases, in order to capture the secular perturbations on
the system. We found that the stable region found in the region 1 for a period of 2 years was slightly decreased in 2000 years. 
Only the most external particles were completely removed, in the interval of 592 years. The stable regions found in region 2, for the period of two years, disappeared in the 
long period. The stability found in region 3 for 2 years does not change along 2000 years. Therefore, we see that the triple 2001 SN263 has a quick scattering of particles, 
i.e., most collisions and ejections happen in the early years.

The stable regions found for 2000 years are an indicative of where satellites or debris could be located in the system. 
Although the time scale of 2000 years is not sufficient to guarantee that, our results restrict the area to be considered in the future research in this direction, 
and guide the spatial mission in the search of  such bodies. On the other hand, we found that the unstable regions occur mostly in a short period of time (2 years), so, 
these regions are least likely to find debris. Thus, thinking in the spatial mission, this is a good region to be investigated, in order to find sets of initial conditions 
that are unstable after 2 years, but that are stable in a shorter time span, enough to accomplish the mission.

According to \cite{b4}, the $J_{2}$ coefficient for Alpha is not a well-constrained value $(J_{2}=0.013 \pm 0.008)$, and 
the present work shows that this coefficient plays an important role in the dynamics of the system.
A change on the $J_{2}$ value is expected to modify such dynamics, as the resonance locations and the effects due to the Kozai mechanism on the inclined cases. 
A study on the stable and unstable
regions as a function of the $J_{2}$ coefficient demand a substantial effort in terms of new simulations and analysis, being thus, the subject of a future work.

\section{Acknowledgments}
This work was funded by INCT - Estudos do Espa\c co, CNPq and FAPESP.  This support is
gratefully acknowledged. We thank Ernesto Vieira Neto for his contribution with the numerical integrator. 
The authors are also grateful to the anonymous referee for all his suggestions.

\renewcommand{\refname}{REFERENCES}

\label{lastpage}


\begin{thebibliography}{27}
\bibitem[\protect\citeauthoryear{}{}]{b}

\bibitem[\protect\citeauthoryear{Amata}{2009}]{b1}AMATA, G.B.; Marco Polo Mission - Executive Summary. 15p. 2009.

\bibitem[\protect\citeauthoryear{Anders}{1964}]{b14} ANDERS,E.,Origin, age and composition of meteorites.,Space Science Review, 3, 574-583, 1964.

\bibitem[\protect\citeauthoryear{D'arrigo}{2003}]{b2}D'ARRIGO, P. The ISHTAR Mission Executive Summary for Publication
on ESA Web Pages. Madrid. 8p. Technical report. 2003.

\bibitem[\protect\citeauthoryear{Becker et al.}{2009}]{b16}BECKER, T. et al.,Physical Modeling of Triple Near-Earth Asteroid 153591 (2001 SN263). American Astronomical Society, DPS meeting 40, 28.06; Bulletin of the American Astronomical Society, Vol. 40, p.437, 2009.

\bibitem[\protect\citeauthoryear{Bottke and Melosh}{1996}]{b17} BOTTKE W. ; H. J. MELOSH,Binary Asteroids and the Formation of Doublet Craters, Icarus,124, p. 372-391, 1996.

\bibitem[\protect\citeauthoryear{Brozovic et al.,}{2009}]{b25} BROZOVIC, M.; BENNER, L. A. M.; NOLAN, M. C, et al. (136617) 1994 CC. IAU Circ., 9053, 2, 2009.

\bibitem[\protect\citeauthoryear{Domingos et al.}{2006}]{b18} DOMINGOS, R. C.; WINTER, O. C.; YOKOYAMA, T. Stable satellites around extrasolar giant planets. MNRAS, V. 373, p.  1227-1234, 2006. 
 
\bibitem[\protect\citeauthoryear{Everhart}{1985}]{b3}EVERHART, E. An efficient integrator that uses Gauss-Radau spacings. In Dynamics of comets: Their origin and evolution, Eds. A. Carusi Carusi and G. B. Valsecchi, D.Reidel Publishing Company (Holanda), p. 185-202, 1985.

 \bibitem[\protect\citeauthoryear{Fang et al.}{2011}]{b4}FANG, J. et al., Orbits of near-earth asteroid triples 2001 SN263 and 1994 CC: properties, origin, and evolution. Astronomical Journal. Volume 141, Issue 5, 2011. 

\bibitem[\protect\citeauthoryear{Fang and Margot}{2012}]{b26} FANG, J.; Margot, J.L. The role of Kozai cycles in Near-Earth binary asteroids. Astronomical Journal, Volume 143, Issue 3, 2012.

 \bibitem[\protect\citeauthoryear{Galvez}{2003}]{b5} GALVEZ,  A. et al., Near Earth Objects Space Mission Preparation: Don Quijote Mission Executive Summary. Madrid. 9p. Technical report. 2003.

 \bibitem[\protect\citeauthoryear{Gladman et al.}{2000}]{b11} GLADMAN,B. et al., The Near Earth Object Population. Icarus, 146, p. 176-189, 2000.

\bibitem[\protect\citeauthoryear{Kozai}{1962}]{b21} KOZAI, Y. Secular perturbation od asteroids with high inclination and eccentricity, The astronomical journal, v. 67, n.9, 
p. 591-598, 1962.

\bibitem[\protect\citeauthoryear{Marchis et al.,}{2005}]{b19} MARCHIS, F,; DESCAMPS, D; HESTROFFER, D.; BERTHIER,J.Discovery of the triple asteroidal system 87 Sylvia.
Nature, V.436, p. 822-824, 2005.

\bibitem[\protect\citeauthoryear{Margot et al.}{2002}]{b23} MARGOT, J. L.; NOLAN, M. C.; BENNER, L. A. M.; OSTRO, S. J.; JURGENS, R. F.; GIORGINI, J. D.; SLADE, M. A.; CAMPBELL, D. B.
Binary Asteroids in the Near-Earth Object Population. Science, V. 296,  no. 5572  p. 1445-1448, 2002.

\bibitem[\protect\citeauthoryear{Morbidelli et al.}{2002}]{b15} MORBIDELLI et al.,Origin and Evolution of Near-Earth Objects.,Asteroids III, p. 409-422, 2002.

\bibitem[\protect\citeauthoryear{Murray and Dermott}{1999}]{b22} MURRAY, D.C.; DERMOTT, S.F. Solar System Dynamics. Cambridge University Press. 1999. 

\bibitem[\protect\citeauthoryear{Nolan et al.,}{2008}]{b6} NOLAN, M.C. et al., Arecibo radar imaging of 2001 SN263: a near-earth triple asteroid system. Asteroids, Comets, Meteors, n. 8258, 2008.

\bibitem[\protect\citeauthoryear{Opik}{1961}]{b12} OPIK,E.J.,The survival of comets and comets material. Astronomical Journal, 66, p. 381-382, 1961.
Astrophysics, 2, 219?262, 1963.

\bibitem[\protect\citeauthoryear{Pravec et al.}{2006}]{b24} PRAVEC, P.; SCHEIRICH, P.; KUSNIRAK, P; et al. Photometric survey of binary near-Earth asteroids.
Icarus, V. 181, Issue 1, p.  63-93, 2006.

\bibitem[\protect\citeauthoryear{Roy}{1988}]{b27} ROY, A.E. Orbital Motion. 3 ed. Institute of Physics Publishing. 1988.

\bibitem[\protect\citeauthoryear{Sears}{2004}]{b7}SEARS,  D. et al.,  The Hera mission: multiple near-earth asteroid
sample return. Advances in space research, v. 34, p. 2270-2275, 2004.

\bibitem[\protect\citeauthoryear{Sukhanov}{2010}]{b8}SUKHANOV, A. et al., The Aster Project: Flight to a Near Earth Asteroid. Cosmic Research, 2010, Vol. 48, No. 5 pp. 443-450.

\bibitem[\protect\citeauthoryear{Wells}{2003}]{b9} WELLS, N. SIMONE NEO Mission Study Executive Summary. Hampshire.11p. Technical report. 2003.

\bibitem[\protect\citeauthoryear{Winter et al.}{2009}]{b20} WINTER, O. C.; BOLDRIN, L. A. G.; VIEIRA NETO, E.; VIEIRA MARTINS, R.; GIULIATTI WINTER, S. M.; GOMES, R. S.; MARCHIS, F.; DESCAMPS, P. 
On the stability of the satellites of asteroid 87 Sylvia, MNRAS, V. 395, Issue 1, p. 218-227, 2009.

 \bibitem[\protect\citeauthoryear{Yoshikawa}{2006}]{b10} YOSHIKAWA,  M. et al., Technologies for
future asteroid exploration: What we learned from hayabusa mission. Spacecraft
Reconnaissance of Asteroid and Comet Interiors, n.3038,2006.

\end{thebibliography}
\end{document}